\documentclass[twocolumn,prd,showpacs,10pt,superscriptaddress]{revtex4}
\usepackage{graphicx}

\def\AA{{\cal A}}
\def\BB{{\cal B}}
\def\CC{{\cal C}}
\def\DD{{\cal D}}
\def\EE{{\cal E}}
\def\LL{{\cal L}}
\def\MM{{\cal M}}
\def\etal{{\em et al.}}
\def\be{\begin{equation}}
\def\ee{\end{equation}}
\def\bea{\begin{eqnarray}}
\def\eea{\end{eqnarray}}
\def\alm{a_{\ell m}}
\def\smax{s_{\rm max}}
\def\lmax{\ell_{\rm max}}
\def\tr{{\rm tr}}
\def\erf{{\rm erf}}
\def\muK{\mu{\rm K}}

\begin{document}

\title{Constraining topology in harmonic space}

\date{\today}

\author{M. Kunz}
\email{Martin.Kunz@physics.unige.ch}
\affiliation{D\'{e}partement de Physique Th\'{e}orique, Universit\'{e} de Gen\`{e}ve,
24 quai Ernest Ansermet, CH-1211 Geneva 4, Switzerland}
\affiliation{Astronomy Centre, SciTech, University of Sussex, Brighton, BN1 9QJ, UK}
\author{N. Aghanim}
\affiliation{Institut d'Astrophysique Spatiale (IAS), B\^{a}timent 121, F-91405, Orsay (France);
Universit\'{e} Paris-Sud 11 et CNRS (UMR 8617)}
\author{L. Cayon}
\affiliation{Department of Physics, Purdue University, 525 Northwestern Ave.,
West Lafayette, IN 47907-2036, USA}
\author{O. Forni}
\affiliation{Institut d'Astrophysique Spatiale (IAS), B\^{a}timent 121, F-91405, Orsay (France);
Universit\'{e} Paris-Sud 11 et CNRS (UMR 8617)}
\author{A. Riazuelo}
\affiliation{Institut d'Astrophysique de Paris, Universit\'e Pierre et Marie Curie,
Paris VI, 98 bis bd Arago, 75014 Paris, France}
\author{J.P. Uzan}
\affiliation{Institut d'Astrophysique de Paris, Universit\'e Pierre et Marie Curie,
Paris VI, 98 bis bd Arago, 75014 Paris, France}

\begin{abstract}
We consider several ways to test for topology directly in harmonic space
by comparing the measured $\alm$ with the expected correlation matrices.
Two tests are of a frequentist nature while we compute the Bayesian
evidence as the third test. Using correlation matrices for cubic and
slab-space tori, we study how these tests behave as a function
of the minimal scale probed and as a function of the size of the
universe. We also 
apply them to different first-year WMAP CMB maps and confirm that the
universe is compatible with being infinitely big for the cases
considered. We argue that there is an information theoretical limit
(given by the Kullback-Leibler divergence) on the size of the
topologies that can be detected.
\end{abstract}

\maketitle

\section{Introduction}

General Relativity has been extremely successful in describing the large-scale
features of our universe. But the global shape of space-time is a quantity that 
is not determined by the local equations of General Relativity. An intriguing 
possibility is therefore that our universe is much smaller than the size of the 
particle horizon today.

In the standard model, the universe is described by a 
Friedmann-Lema\^\i tre-Robertson-Walker (FLRW) type metric which is 
homogeneous and isotropic. If the topology
of the universe is not trivial, then we are dealing with a quotient space
$X/\Gamma$ where $X$ is one of the usual simply connected FLRW spaces (spherical,
Euclidean or hyperbolic) and $\Gamma$ is a discrete and fixed-point free symmetry
group that describes the topology. This construction does not affect local physics
but changes the boundary conditions (see eg.~\cite{rep1,rep2} and references
therein). 

This could potentially explain some of the anomalies found in the first-year WMAP data.
For example, the perturbations of the cosmic fluids need to be invariant
under $\Gamma$. Therefore  the largest
wavelength of the fluctuations in the CMB cannot exceed the size of the
universe, and so the suppression (and maybe the strange alignment) of the
lowest CMB multipoles might be due to a non-trivial topology 
\cite{cl1,cl2,cl3,cl4,align,align2}. Additionally,
the last scattering surface can wrap around the universe. In this case
we receive CMB photons, which originated at the same physical location on the 
last scattering surface, from different directions. Observationally
this would appear as matched
(correlated) circles in the CMB \cite{circle}.
An analysis by Cornish et al of the first-year WMAP maps
based on a search for matching circles has not found any evidence for a
non-trivial topology \cite{cornish2}. However, it is difficult to quantify the
probability of missing matching circles, and other groups have claimed
a tentative detection of circles at scales not probed by Cornish et al
(see e.g. \cite{roukema,luminet}). In this paper we study a different approach which can
in principle yield both an optimal test as well as a rigorous assessment
of the fundamental detection power of the CMB for a cosmic topology.

Instead of working directly with the observed map of CMB temperature
fluctuations, we expand the map in terms of spherical harmonics,
\be
T(x) = \sum_{\ell, m} \alm Y_{\ell m}(x),
\ee
where $x$ are the pixels. Both the pixels and the expansion coefficients $\alm$
are random variables. In the simplest models of the early universe, they are
to a good approximation Gaussian random variables, an assumption that we will
make throughout this paper. Their $n$-point correlation functions are then
completely determined by the two-point correlation function.
The homogeneity and isotropy of the simply-connected 
FLRW universe additionally requires the two-point correlation of the $\alm$ 
to be diagonal,
\be
\langle \alm a_{\ell'm'}^* \rangle = C_\ell \delta_{\ell \ell'} \delta_{m m'} .
\ee
The symmetry group $\Gamma$ will introduce preferred directions, which will
break global isotropy. This in turn induces correlations between off-diagonal
elements of the two-point correlation matrix. In this paper we study methods
to find such off-diagonal correlations. Such a test is complementary to the
matched-circle test of \cite{circle,cornish2}, and if the initial fluctuations are
Gaussian then it can use {\em all} the information present in the CMB
maps and so lead to optimal constraints on the size of the universe.
Investigating the amount of information introduced into the two-point
correlation matrix by a given topology allows us to decide from an
information theoretical standpoint whether the CMB will ever be able
to constrain that topology.

We will use the following notation: We often
combine the $\ell$ and $m$ indices to a single index 
$s\equiv\ell(\ell+1)+m$ and mix
both notations frequently. The noisy correlation
matrix given by the data is $\AA_{ss'} \equiv a_s a_{s'}^*$.
We will write the correlation matrix
which defines a given topology as $\BB_{ss'}$. This is the expectation
value of the two-point correlation function for $a_s$ that describe
a universe with that topology.

All the simulations in this paper are based on a flat $\Lambda$CDM model
with $\Omega_\Lambda=0.7$, a Hubble parameter of $h=0.67$, a Harrison-Zel'dovich
initial power spectrum ($n_S=1$) and a baryon density of $\Omega_b h^2=0.019$,
as described in \cite{riaz1,riaz2}. With this choice of cosmological parameters
we find a Hubble radius $1/H_0 \approx 4.8$Gpc while the radius of the particle
horizon is $R_h \approx 15.6$Gpc. 
We will denote a toroidal topology as T[X,Y,Z] where X, Y and Z are the
sizes of the fundamental domains, in units of the Hubble radius.
As an example, T[4,4,4] is a cubic torus of size $(19.3\textrm{Gpc})^3$.
The volume of such a torus is nearly half that of the observable universe.
The diameter of the particle horizon is about $6.5/H_0$.
But we should note that there are non-zero off-diagonal terms in
$\BB_{ss'}$ even for universes that are slightly larger than the
particle horizon. 

We have a range of correlation matrices at our disposal so far.
Two of them are cubic tori with sizes $2/H_0$ (T[2,2,2]) and $4/H_0$
(T[4,4,4]). For these two we have the correlation matrices up to 
$\lmax=60$ (corresponding to $\smax=3720$). We also have two families
of slab spaces. The first one, T[X,X,1], has one very small direction
of size $1/H_0$. The second one, T[15,15,X], has two large directions that
are effectively infinite. Both groups include all tori with $X=1,2,\ldots,15$,
and we know their correlations matrices up to $\lmax=16$ (or $\smax=288$).
The correlation matrices analysed in this paper do not contain the
integrated Sachs-Wolfe contributions (cf discussion in section \ref{sec:isw}).

This paper is organised as follows: We start out by matching the
measured correlations to a given correlation matrix.
We then show that a similar power to distinguish between
different correlation matrices can be achieved by using the
likelihood. In general we do not know the relative orientation
of the map and the correlation matrix, and we discuss how to
deal with this issue next. We then present a first set of
results from this analysis, before embarking on a simplified
analysis of the WMAP CMB data and toroidal topologies.

So far the methods were all of a frequentist nature. Using the likelihood
we can also study the evidence for a given topology, which is the
Bayesian approach to model selection. We then talk about the
issues that we neglected in this paper, and finish with conclusions.

The appendices look in more detail at how the correlation and the
likelihood method differ, and how their underlying structure can be
used to define ``optimal'' estimators. We also discuss how selecting
an extremum over all orientations can be linked to extreme value distributions, 
which allows us to derive probability distribution functions that can be 
fitted to the data for
quantifying confidence levels. We finally consider a distance function
on covariance matrices, motivated by the Bayesian evidence discussion,
and study its application to the comparison between different topologies.

\section{Detecting Correlations}

A priori it is very simple to check whether there are significant off-diagonal
terms present in the two-point correlation matrix: One just looks at terms
with $\ell\neq\ell'$ and/or $m\neq m'$. But the variance of the
$\alm$ is too large as we can observe only a single universe. When computing 
the $C_\ell$ we average over all directions
$m$. This averaging then leads to a cosmic variance that behaves like $1/\sqrt{\ell}$.
But now we have to consider each element of the correlation matrix separately,
leading to a cosmic variance of order $1$ for each element. The matrix is
therefore very noisy and we need to ``dig out'' the topological signal from 
the noise. Furthermore, if we detect the presence of significant off-diagonal
correlations, we still need to verify that they are due to a non-trivial
topology and not to some other mechanism that breaks isotropy.

A natural approach to the problem is then to use the expected correlation
matrix for a given topology as a kind of filter.
To this end we compute a correlation amplitude $\lambda$
which describes how close two matrices are. We do this by minimising
\be
\chi^2[\lambda] = \sum_{s s'} 
\left(\AA_{s s'} - \lambda \BB_{s s'}\right)^2 \label{eq:chi2}
\ee
where $\AA_{s s'}=a_s a_{s'}^*$ is the correlation matrix estimated from
the data and $\BB_{s s'}$ the one which contains the topology that we want to test. 
For a good fit we expect to find $\lambda\approx 1$
while for a bad fit $\lambda\approx0$.

We can easily solve $d\chi^2/d\lambda=0$ and find that
\be
\lambda = \frac{\sum_{s s'} \AA_{s s'} \BB_{s s'}}{
  \sum_{s s'} (\BB_{s s'})^2}  \label{eq:corpar}
\ee
minimises Eq.~(\ref{eq:chi2}). 
As we know that we will have to compare our method against maps
from an infinite universe with the same power spectrum, we do not sum
over the diagonal $s=s'$ (which corresponds to $\ell=\ell'$ and
$m=m'$) to improve the signal to noise. This corresponds to
replacing the correlation matrix through
$\BB\rightarrow\BB-\DD$ where $\DD$ is a diagonal matrix with
the power spectrum on the diagonal. If the power spectrum
is constant so that $\DD = C\times1$ then
the eigenvectors of the new correlation matrix are the same as
those of the original one, and the eigenvalues are replaced
by $\epsilon^{(i)}\rightarrow\epsilon^{(i)} - C$. In this
case they will no longer be positive.

We could also introduce a covariance matrix in Eq.~(\ref{eq:chi2}). In
the presence of noise this may be useful. In this study we will assume
throughout an idealised noise-free and full-sky experiment for simplicity.
At any rate the WMAP data will be cosmic variance dominated at the low
$\ell$ that we consider here, see section \ref{sec:noise}. 
Neglecting the noise contribution, the covariance matrix is 
$C_{qq'}=\langle\BB_q\BB_{q'}\rangle$ where $q=\{s,s'\}$. 
But as the correlation matrices
are already expectation values, we end up with a matrix that has a
single non-zero eigenvalue $\epsilon = \sum_q \BB_q^2$. If we invert
this singular matrix with the singular value decomposition (SVD) method 
(setting the inverse of the zero eigenvalues to zero) and minimise the 
resulting expression for
the $\chi^2$, we find again Eq.~(\ref{eq:corpar}).

It is straightforward to compute the expectation value and variance of
the $\lambda$ function for two important cases. In the first case the
universe is infinite, so that the spherical harmonics $\alm$ are
characterised by the usual two-point function,
\be
\langle \alm a_{\ell'm'}^*\rangle_\infty = C_\ell \delta_{\ell\ell'}\delta_{mm'} .
\label{eq:infcor}
\ee
In the second case the universe has indeed the topology described
by the correlation matrix $\BB$ against which we test the $\alm$.
In this
case the two point function of the spherical harmonics is given by
\be
\langle \alm a_{\ell'm'}^*\rangle_\BB = \BB_{ss'}
\ee
In both
cases the spherical harmonics obey a Gaussian statistics and the higher
$n$-point functions are uniquely determined by the two-point function
via Wicks theorem.

Let us first define the auto-correlation $U=\sum_{ss'} |\BB_{ss'}|^2$.
We remind the reader that such sums in this section exclude the diagonal
terms $s=s'$ except where specifically mentioned.
For an infinite universe, we notice that if we sum only over the
non-diagonal elements $s \ne s'$ then,
since $\langle a_s a_{s'}^* \rangle_\infty = C_s \delta_{ss'}$ the
expectation value of lambda is zero, $\langle \lambda \rangle_\infty =
0$. Else,
\be
\langle \lambda \rangle_\infty = \frac{1}{U} \sum_s C_s \BB_{ss} .
\ee
If the map was whitened (see below), 
then $\langle \lambda \rangle_\infty = \tr(\BB)/U$.

For a finite universe,
\be
\langle\lambda\rangle_\BB = 1
\ee
independently if we sum over the diagonal elements or not, as we just recover
the auto-correlation in the numerator. Of course the auto-correlation value
depends on the summation convention.

For the variance, in the case of an infinite universe, we find
\be
\sigma_\infty^2
\equiv\langle \lambda^2 \rangle_\infty - \langle \lambda \rangle_\infty^2
= \frac{2}{U^2} \sum_{ss'} C_s C_{s'} |\BB_{ss'}|^2 .
\ee
The summation depends again if we keep the diagonal elements or
not. For a whitened map, the result simplifies to $\sigma_\infty^2 = 2/U$.
In a finite universe,
\be
\sigma^2_\BB = \frac{2}{U^2} \tr\left(\BB\BB^*\BB\BB^*\right) ,
\ee
however now we need to be more careful if we discard the diagonal elements,
as then
\be
\sigma^2_\BB = \frac{2}{U^2} \sum_{s1\neq s2,s3\neq s4}
\left(\BB_{s1s2}\BB^*_{s2s3}\BB_{s3s4}\BB^*_{s4s1}\right)
\label{eq:corr_A_error}
\ee
Table \ref{tab:lambda} shows the expectation values of variances
for a selection of topologies, computed with these formulas. It may
be surprising that the variance of $\lambda$ for an infinite universe
depends on the test-topology. However, Eq.~(\ref{eq:corpar}) depends
on $\BB$ even if the $\alm$ do not. The variance is a measure of how
different $\BB$ is from the diagonal ``correlation matrix'' of an
infinite universe, Eq.~(\ref{eq:infcor}). The larger the difference,
the smaller the variance of $\lambda$, as the random off-diagonal
correlations present in the $\alm$ are less likely to match those
of the test-matrix $\BB$. The value of $\lmax$ in the table was
chosen basically arbitrarily, we will discuss later how it influences
the measurements. We have also introduced a ``signal to noise ratio'' S/N
which is the difference of the expectation values, divided by the
errors added in quadrature,
\be
S/N(\BB,X) = 
\frac{|\langle X \rangle_\infty - \langle X\rangle_\BB|}
{\sqrt{\sigma(X)^2_\infty+\sigma(X)^2_\BB}} .
\label{eq:sn}
\ee
Here $X$ is the estimator used. This gives only a rough indication
of the true statistical significance with which a universe with
the given topology can be distinguished from an infinite universe.
As the distribution of $\lambda$ and $\chi^2$ are not exactly Gaussian,
S/N is not exactly measured in units of standard deviations. However,
it is sufficient to compare the different methods and to 
illustrate how well different topologies can be detected. For
precise statistical results we fit the full distribution, see
appendix \ref{app:rot}.

\begin{table}[ht]
\begin{tabular}{|l|cccc|}
\hline
topology & $\ell_{\rm max}$ & $\lambda_\infty$ & $\lambda_\BB$ & S/N [$\sigma$] \\
\hline
T[2,2,2]     &       60         & $0\pm0.017$  & $1\pm0.102$  & $9.7$  \\
T[4,4,4]     &       60         & $0\pm0.046$  & $1\pm0.082$  & $10.6$  \\
T[2,2,2]     &       16         & $0\pm0.03$ & $1\pm0.34$   &  $2.9$  \\
T[4,4,4]     &       16         & $0\pm0.09$ & $1\pm0.22$   &  $4.2$  \\
T[6,6,1]     &       16         & $0\pm0.08$ & $1\pm0.33$   &  $2.9$  \\
T[15,15,6]  &       16         & $0\pm0.51$ & $1\pm0.59$   &  $1.3$  \\
\hline
\end{tabular}
\caption{Comparison of the mean and standard deviation of $\lambda$ for
different topologies and different $\ell_{\rm max}$, normalised with the
true power spectrum. The S/N value is given by Eq.~(\ref{eq:sn}).
\label{tab:lambda}}
\end{table}

The power spectrum $C_\ell$ depends of course on the cosmological
parameters. To minimise this potential
problem we normalise the correlation matrices either by the diagonal
$C_s\equiv \langle a_s a^*_s \rangle$ or by the usual orientation-
averaged power spectrum
\be
C_\ell = \frac{1}{2\ell+1} \sum_m |\alm|^2 ,
\ee
via the prescription 
\be
\BB_{ss'} \rightarrow \frac{\BB_{ss'}}{\sqrt{C_s C_{s'}}} .
\ee
This is often called ``whitening'', and it serves to enforce the
same (white noise) power spectrum in both the template and the
model being tested. {\em After} applying this normalisation the
power spectrum is just $C_s=1$. We apply the same normalisation
to the $\alm$. As we will not in general know their ``true'' input
power spectrum,
we use the one recovered from the $\alm$ themselves. 
As can be seen in table \ref{tab:lambdadiv},
the division by the recovered power spectrum greatly reduces the
variance of $\lambda$ and so improves the detection power for the
different topologies. Contrary to table \ref{tab:lambda} we could
not compute the numbers analytically and have estimated them from
$10^4$ random realisations each of maps with the trivial topology
and the $\BB$ topology.

\begin{table}[ht]
\begin{tabular}{|l|cccc|}
\hline
topology & $\ell_{\rm max}$ & $\lambda_\infty$ & $\lambda_\BB$ & S/N [$\sigma$] \\
\hline
T[2,2,2]     &       60         & $0\pm0.015$ & $0.973\pm0.030$ & $29.0$  \\
T[4,4,4]     &       60         & $0\pm0.051$ & $0.976\pm0.044$ & $14.5$  \\
T[2,2,2]     &       16         & $0\pm0.032$ & $0.924\pm0.100$ & $8.8$  \\
T[4,4,4]     &       16         & $0\pm0.091$ & $0.948\pm0.100$ & $7.0$  \\
T[6,6,1]     &       16         & $0\pm0.083$ & $0.894\pm0.200$ & $4.1$  \\
T[15,15,6]  &       16         & $0\pm0.534$ & $0.971\pm0.553$ & $1.3$  \\
\hline
\end{tabular}
\caption{Comparison of the mean and standard deviation of $\lambda$ for
different topologies and different $\ell_{\rm max}$, normalised with the
power spectrum {\em estimated independently for each realisation}. As we
see, the signal to noise ratio is improved considerably.
\label{tab:lambdadiv}}
\end{table}

For an infinite universe $C_s$ is independent of $m$ and it does
not matter whether we divide by $C_s$ or $C_\ell$.
For non-trivial topologies this is not
the case as additional correlations are induced in different
$m$ modes. For this reason, the division by the $m$-averaged $C_\ell$
tends to lead to somewhat stronger constraints.
Of course we lose the information encoded in the power spectrum,
like the suppression of fluctuations with wavelengths larger than
the size of the universe. However, we feel that the improved
stability to mis-estimates of the power spectrum and the reduced
dependence on the cosmological parameters is worth the trade-off.

The numerical evaluation of Eq.~(\ref{eq:corpar}) requires a double
sum over $\smax=\lmax(\lmax+2)$ matrix coefficients. It scales
therefore as $\lmax^4$. But the correlation matrix of an infinite
universe is diagonal, so that we only need to perform a single
sum. It should therefore be possible to reduce the work for matrices
that are close to being diagonal, ie.~for universes with a very large
compactification scale. A possibility is to decompose the
correlation matrix into a sum over eigenvalues and
eigenvectors. We can then only retain the most important eigenvectors.
As the correlation
matrix is also a covariance matrix, this is somewhat analogous to
principal component analysis or the Karhunen-Loeve transform.
For a correlation matrix $\BB$ we will write the decomposition as
\be
\BB_{ss'} = \sum_i \epsilon^{(i)} v^{(i)}_s v^{(i)*}_{s'}
          = \sum_i b^{(i)}_s b^{(i)*}_{s'} .
\label{eq:evec}
\ee
The $\epsilon^{(i)}$ are the eigenvalues
of the matrix $\BB$ and they are real and positive as the matrix is hermitian and positive. This
allows us to define effective spherical harmonics 
$b^{(i)}_s\equiv\sqrt{\epsilon^{(i)}}v^{(i)}_s$, which have, for example,
the same properties under rotation as the usual $\alm$.

\section{Using the likelihood\label{sec:like}}

Instead of considering the correlation between the recovered and
the theoretical matrix, we can think of the two-point
correlation matrix as the covariance matrix of the $\alm$. Then
we may ask the question, what is the probability of a
covariance matrix $\CC$ given the measured $\alm$. This can be
answered using Bayesian statistics. 

In a first step we need to construct the likelihood function.
The probability distribution for a Gaussian random variable $x$
with variance $\sigma^2$ and zero expectation value is
\be
p(x|\sigma) = \frac{1}{\sqrt{2\pi}\sigma} e^{-\frac{x^2}{2\sigma^2}}
\ee
If we  assume that we measure $x$ and want to know $\sigma$, then 
the likelihood function for finding a certain $x$ is given by 
$\LL(\sigma) \equiv p(x|\sigma)$. We write the likelihood as a function
of the variance, as this is the model parameter that we are interested
in.

For many
independent variables, the probability distribution is the
product, which leads to a sum in the exponent. In the case of
the $\alm$, the random variables are
not independent but are distributed according to a
multivariate Gaussian distribution with a covariance matrix
$\CC$. The likelihood function then is
\be
p(\alm|\CC) =
\LL(\CC) \propto \frac{1}{\sqrt{|\CC|}} \exp\left\{-\frac{1}{2}
\sum_{s,s'} a^*_s  \CC^{-1}_{ss'} a_{s'} \right\} ,
\ee
where $|\CC|$ is the determinant of the matrix $\CC$. The covariance
matrix is given by the two-point correlation matrix, and
$\langle a_s \rangle=0$. Any further model assumptions are
implicitly included in the choice of $\CC$.
Using Bayes law we can invert the probability to find
\be
p(\CC|\alm) = \frac{p(\alm|\CC) p(\CC)}{p(\alm)} .
\ee
The probability in the denominator is a normalisation constant, while
$p(\CC)$ is the prior probability of a given topology encoded by
$\CC$. We will assume that we have no prior information
about the topology of the universe, so that this is a constant as
well. In this case $p(\CC|\alm)\propto\LL(\CC)$, ie. we can use the
likelihood function to estimate the probability of a topology
given a set of $\alm$. For our purpose, the covariance matrix is
just given by the correlation matrix $\BB$. In general, one may have
to add noise to it, and maybe introduce a sky cut.

Generally it is preferable to consider the logarithm of the likelihood,
$\log(\LL) = -1/2(\log(|\BB|)+\chi^2)+{\rm const.}$ where we have
defined
\be
\chi^2 = \sum_{s,s'} a^*_s  \BB^{-1}_{ss'} a_{s'} .
\ee

We notice that there is a potential issue with the normalisation of
the input model: If $a_s \rightarrow 0$ then $\chi^2\rightarrow0$ --
generally any model whose $a_s$ lead to a bad fit (high $\chi^2$) could
be renormalised until a reasonable likelihood is obtained. It is therefore
required to fix the overall normalisation, and we will do this
by using the whitened $a_s$, in which case the normalisation
is fixed by $\sum_s |a_s|^2=1$.

For the two special cases, the infinite universe and $\alm$ distributed
according to $\BB$, we can compute expectation value and variance.
For the general case we will write
$\langle a_s a_{s'}\rangle = \AA_{ss'}$. Then
\be
\langle \chi^2 \rangle = \sum_{ss'} \langle a_s^* a_{s'}\rangle \BB_{ss'}^{-1}
= \tr(\AA\BB^{-1}) ,
\ee
where we have used the hermeticity of the correlation matrices. The
two special cases are
\bea
\langle \chi^2 \rangle_\infty &=& \sum_s C_s \BB_{ss}^{-1} \\ 
\langle \chi^2 \rangle_\BB &=& \tr(1) = \smax
\eea

As the $\alm$ are Gaussian random variables, we expect to find
that $\chi^2$ is distributed with a $\chi^2$-like distribution. 
The general expression is rather cumbersome, but for the two
special cases we find
\be
\sigma^2_\BB \equiv 
\langle (\chi^2)^2 \rangle_\BB - \langle \chi^2 \rangle^2_\BB
= 2 \smax
\ee
and
\be
\sigma^2_\infty = 2 \sum_{ss'} C_s C_{s'} |\BB^{-1}_{ss'}|^2 .
\ee
We list in table \ref{tab:chi2} some examples, together with the number of
standard deviations that the two expectation values lie apart. 

\begin{table}[ht]
\begin{tabular}{|l|cccc|}
\hline
topology & $\ell_{\rm max}$ & $\chi^2_\infty$ & $\chi^2_\BB$ & S/N [$\sigma$] \\
\hline
T[2,2,2]     &       60         & $37168\pm2373$  & $3720\pm86$  & $14.1$  \\
T[4,4,4]     &       60         & $14656\pm1517$  & $3720\pm86$  &  $7.2$  \\
T[2,2,2]     &       16         & $5608\pm738$    & $288\pm24$   &  $7.2$  \\
T[4,4,4]     &       16         & $1802\pm300$    & $288\pm24$   &  $5.0$  \\
T[6,6,1]     &       16         & $20781\pm7103$  & $288\pm24$   &  $2.9$  \\
T[15,15,6]  &       16         & $309\pm28$      & $288\pm24$   &  $0.6$  \\
\hline
\end{tabular}
\caption{Same as table \ref{tab:lambda} for $\chi^2$.
\label{tab:chi2}}
\end{table}

In these computations, as in the corresponding ones for the
correlation coefficient, 
we have assumed that we normalise the observed $\alm$ by the
``true'' power spectrum (or diagonal). However, we do not know what it is.
If we instead normalise them by the estimated one (which is different for
each realisation), we change the statistics. It is now no longer Gaussian.
Table \ref{tab:chi2div} reproduces the previous
one, but now for this scenario. We estimated the numbers from $10^4$
numerical realisations for each topology. Again the detection power
increases considerably.
\begin{table}[ht]
\begin{tabular}{|l|cccc|}
\hline
topology & $\ell_{\rm max}$ & $\chi^2_\infty$ & $\chi^2_\BB$ & S/N [$\sigma$] \\
\hline
T[2,2,2]     &       60         & $37366\pm1123$ & $4655\pm438$ & $27.1$ \\
T[4,4,4]     &       60         & $14932\pm1157$ & $4027\pm162$ &  $9.3$ \\
T[2,2,2]     &       16         & $5690\pm477$   & $474\pm131$  & $10.5$ \\
T[4,4,4]     &       16         & $1841\pm196$   & $335\pm48$   &  $7.5$ \\
T[6,6,1]     &       16         & $21093\pm5645$ & $786\pm557$  &  $3.6$ \\
T[15,15,6]  &       16         & $309\pm10$     & $289\pm5$    &  $1.8$ \\
\hline
\end{tabular}
\caption{Same as table \ref{tab:lambdadiv} for $\chi^2$.
\label{tab:chi2div}}
\end{table}

In appendix \ref{app:opt} we compare the structure of the correlation
estimator to the likelihood $\chi^2$. We find that for many cases the
$\chi^2$ has minimal variance.

\section{Rotating the map into position}

\begin{figure}[ht]
\begin{center}
\includegraphics[width=70mm]{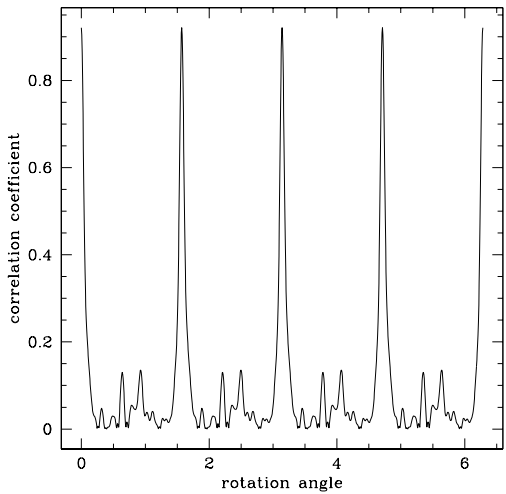} \\
\caption{\label{fig:rot} Behaviour of the correlation coefficient
$\lambda$ under a rotation around the z-axis. The signal is maximal
only for very well-defined alignments. We used a T[2,2,2] correlation
matrix and $\alm$ derived from a T[2,2,2] topology.}
\end{center}
\end{figure}

The situation discussed so far is somewhat misleading: 
Nature is rather unlikely
to align the topology of the universe with our coordinate system.
The correlation matrices are not invariant under rotations,
as rotations mix $\alm$ with different $m$. 
To parametrise the rotations we use the three Euler angles
$\alpha$, $\beta$ and $\gamma$ which describe three subsequent
rotations around the $z$, the $y$ and again the $z$ axis. The
first and last rotation just lead to a phase change.
The rotation around the y-axis couples different $m$ and is
given by Wigner rotation matrices $d^\ell_{mm'}$,
\be
\alm \rightarrow \sum_{m'} e^{-i(m\alpha+m'\gamma)}
d^\ell_{mm'}(\beta) a_{\ell m'} .
\ee
Together, the three rotations can describe any element of the
rotation group of order $\ell$. We use the relations given
in \cite{choi} to compute the rotation matrices.
Figure \ref{fig:rot} shows an example
where we plot $\lambda$ while rotating the $\alm$ azimuthally.
The figure represents the case for $\lmax=60$, for lower values
of $\lmax$ the peaks are less sharp and there is less sub-structure.
The same is true for the $\chi^2$, while the peaks for likelihood,
which is proportional to $\exp(-\chi^2/2)$, are even much narrower.

\begin{figure}[ht]
\begin{center}
\includegraphics[width=45mm,angle=90]{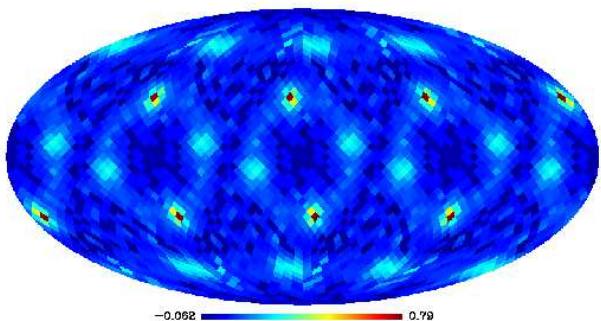} \\
\caption{\label{fig:map} The maximal correlation coefficient for the
  case of a universe with T[2,2,2] orientation. The sharp, high peaks
  correspond to the correct orientation of the map with respect to the
  template.}
\end{center}
\end{figure}

We can therefore not avoid probing all possible rotations, either
by computing the average or by taking
the maximum/minimum of our estimator over all orientations.
Possibly the most straightforward approach is to try many random rotations 
\cite{graca}. This is simple to program and uses automatically any 
symmetries present in the
template. But due to the precision needed to find the best
alignment for some templates, we found that we need in excess of
$10^6$ rotations to get correct results for $\lmax=60$.
We can on the other hand probe
systematically all orientations, for example with the total convolution
method \cite{totconv}. In this approach, the rotations with the three
Euler angles are replaced by a three-dimensional FFT. This speeds the
procedure up a by a large factor. However, we found that
we may nonetheless miss the best-fit peaks which can be very sharp
(see Figs.~\ref{fig:rot} and \ref{fig:map}). 

If we limit ourselves to finding the maximum/minimum efficiently,
then we can also start with a random rotation and search for a
local extremum nearby. We then repeat the procedure for different
random starting locations until we have found a stable global maximum
(for example, eight times the same global maximum). This is the safest
method, and can be relatively fast depending on the topology.

Computing the average is therefore quite difficult and slow.
We also found that using the maximum or minimum results in a much
stronger detection than using the average, at least for the $\lambda$
and $\chi^2$ estimator. It is possible to improve the average by
using the likelihood which is proportional to $\exp(-\chi^2/2)$.
This decreases the weight of the ``wrong'' orientations exponentially.
However, it makes the average even harder to compute. Furthermore,
it lends itself readily to a Bayesian interpretation which is quite
different from the frequentist approach followed so far. For this
reason we will consider only the maximum/minimum approach here and
defer the discussion of the average likelihood to section \ref{sec:evidence}.
We also note that it makes no difference if we consider the
$\chi^2$ estimator or the likelihood when using the extremum
over orientations. The exponential function is monotonic and so
the maximum or minimum point will not change under it (except that
the minimum of the $\chi^2$ will turn into a maximum of the likelihood 
and vice versa). For the
same reason, it does not change the statistical weight. If $99$
realisations of model $A$ have a lower $\chi^2$ than any of model
$B$, then those $99$ realisations will have a higher likelihood
as well.

A drawback of using the extremum over all rotations is that we do not
know the resulting distribution function. In general
we have to compute a large number of test-cases to obtain the distribution,
but this is very time-consuming and for high $\lmax$ computing more
than a few hundred realisations becomes prohibitive, at least on a
single processor. 
Instead we can find a good approximation to the new distribution
by assuming
that each rotation leads to a new independent Gaussian distribution.
If there are $N$ independent rotations 
then we
need to know the distribution of the maximal value of $N$ draws from
a Gaussian distribution. This leads to an extreme value distribution,
and exact results are known only for $N<6$. However, for very large
$N$, the distribution should converge to one of three limiting cases, 
analogously to the central limit theorem (see eg.~\cite{ext_val}).
If we fit these distributions
to the numerical results then we can obtain confidence limits with
a reasonable amount of cpu-time. We discuss this in more detail in
appendix \ref{app:rot}.

\begin{table}[ht]
\begin{tabular}{|l|cccc|}
\hline
topology & $\ell_{\rm max}$ & $\chi^2_\infty$ & $\chi^2_\BB$ & S/N [$\sigma$] \\
\hline
T[2,2,2]     &       60         & $33237\pm586$  & $4588\pm382$ &  $41$ \\
T[4,4,4]     &       60         & $11146\pm438$  & $4057\pm204$ &  $14$ \\
T[2,2,2]     &       16         & $4062\pm172$   & $469\pm172$  &  $17$ \\
T[4,4,4]     &       16         & $1180\pm73$    & $350\pm47$   &  $10$ \\
T[6,6,1]     &       16         & $7719\pm1125$  & $675\pm370$  &  $6$ \\
T[15,15,6]  &       16         & $287\pm2.1$    & $285\pm2.5$  &  $0.6$ \\
\hline
\end{tabular}
\caption{Comparison of the mean and standard deviation of the $\chi^2$ for
different topologies and different $\ell_{\rm max}$, normalised with the
power spectrum and minimised over rotations.
\label{tab:chi2rot}}
\end{table}

We compare in tables \ref{tab:chi2rot} and \ref{tab:lambdarot} the
minimal $\chi^2$ and maximal $\lambda$ values respectively, taken over
all possible orientations. We also quote the resulting S/N value. We
notice that especially the $\chi^2$ estimator gains in sensitivity.
This seems rather surprising, as the distance between the estimator
values of an infinite and a finite universe will in general decrease
when taking the extremum. However, we also notice that the variance
is dramatically decreased, which in turn leads to the even higher
detection power.

The reduction of the variance, especially for the infinite universe case,
is easy to understand. In table \ref{tab:lambdadiv} and \ref{tab:chi2div} 
we use the best-fit
alignment for the maps of a finite universe. But the maps with the trivial
topology are always randomly aligned (being statistically isotropic). The
variance for the infinite universe maps contains therefore an effective
``random orientation'' contribution. Taking the extremum over all orientations
eliminates this contribution. As the infinite universe variance dominates
strongly in the case of the $\chi^2$ estimator, we find that this estimator
benefits more from the reduction of the variance.

\begin{table}[ht]
\begin{tabular}{|l|cccc|}
\hline
topology & $\ell_{\rm max}$ & $\lambda_\infty$ & $\lambda_\BB$ & S/N [$\sigma$] \\
\hline
T[2,2,2]     &       60         & $0.08\pm0.01$ & $0.98\pm0.03$ & $28$  \\
T[4,4,4]     &       60         & $0.21\pm0.02$ & $0.98\pm0.05$ & $14$  \\
T[2,2,2]     &       16         & $0.16\pm0.02$ & $0.95\pm0.08$ & $10$  \\
T[4,4,4]     &       16         & $0.38\pm0.05$ & $0.98\pm0.09$ & $6$  \\
T[6,6,1]     &       16         & $0.35\pm0.05$ & $0.94\pm0.19$ & $3$  \\
T[15,15,6]  &       16         & $1.84\pm0.25$ & $1.86\pm0.27$ & $0$  \\
\hline
\end{tabular}
\caption{Same as table \ref{tab:chi2rot} for $\lambda$.
\label{tab:lambdarot}}
\end{table}

As a final point, we notice that the maximised value of $\lambda$ for
the T[15,15,6] topology in table \ref{tab:lambdarot} is larger than $1$.
This is a sign that we cannot detect that topology. The fluctuations
are so large that they completely overwhelm the signal. After
maximising over orientations we end up with $\lambda > 1$.

\section{Discussion of general results\label{sec:res}}

\subsection{What angular resolution is required?}

Is it better to test the maps to arbitrarily high $\lmax$, or to 
use a lower resolution? One important consideration is the
amount of work (and thus of time) needed to evaluate the
estimator. For both estimators we need to sum over $s$ and
$s'$. This means that the required number of operations scales
like $\lmax^4$. The matrix inversion required for the likelihood
evaluation scales like $\lmax^6$. However, for two
reasons it is normally not the limiting factor. Firstly, as discussed
in the previous section, we still need to average over directions.
To do that we only need to invert the matrix once at the start,
not for every evaluation. But we need to evaluate the likelihood
for each orientation, and the number of the required rotations scales
roughly like $\lmax^2$. We therefore end up with a $\lmax^6$ scaling
at any rate. Secondly, the most time consuming
procedure is the estimation of the variance using simulated
maps, and again we only need to invert the matrix once as it
stays the same. $\lmax^6$ is a rather steep growth,
and it is certainly preferable to use the smallest matrices
that guarantee a detection.

On the other hand, does the detection always improve with growing
$\lmax$? Let us have a look at the correlation estimator, in the
case of a whitened map. Clearly $\sigma_\infty^2 = 2/U$ can only
decrease as long as there are {\em any} off-diagonal elements
in the correlation matrix. But this is not the dominant error.
However, we expect that the main contribution to
Eq.~(\ref{eq:corr_A_error}) is due to the remaining diagonal
entries $s_2=s_3$ and $s_1=s_4$. This term of the sum
is equal to the auto-correlation $U$ and so contributes the same
error as $\sigma_\infty^2$. As the signature of the topology
becomes very weak, we expect that the two errors become
comparable, but are still decreasing functions of $\lmax$.

\begin{figure}[ht]
\begin{center}
\includegraphics[width=70mm]{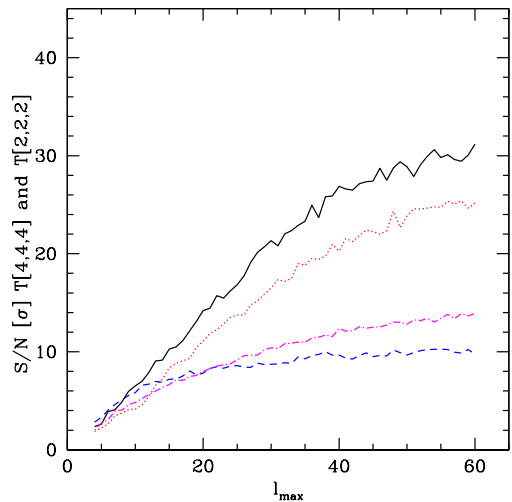} \\
\caption{\label{fig:lscal_norot} Detection significance assuming that
we know the correct orientation. The topologies were T[2,2,2] (solid black
and dotted red line) and T[4,4,4] (dashed blue and dot-dashed magenta line).
The estimators were respectively the correlation amplitude $\lambda$
(dotted red and dot-dashed magenta line) and the likelihood $\chi^2$
(solid black and dashed blue line).}
\end{center}
\end{figure}

\begin{figure}[ht]
\begin{center}
\includegraphics[width=70mm]{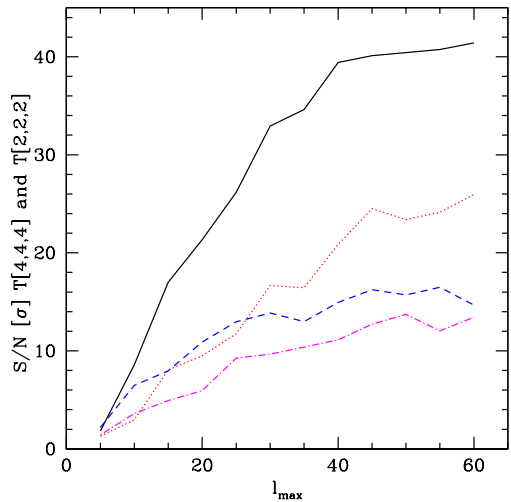} \\
\caption{\label{fig:lscal_withrot} Detection significance when maximising
over all orientations. The topologies were T[2,2,2] (solid black
and dotted red line) and T[4,4,4] (dashed blue and dot-dashed magenta line).
The estimators were respectively the correlation amplitude $\lambda$
(dotted red and dot-dashed magenta line) and the likelihood $\chi^2$
(solid black and dashed blue line).}
\end{center}
\end{figure}

We compare in Figs.~\ref{fig:lscal_norot} and
\ref{fig:lscal_withrot} the scaling of $S/N(T[2,2,2])$ and 
$S/N(T[4,4,4])$ respectively, for the correlation estimator 
(red dotted / magenta dash-dotted) and
the likelihood method (black solid / blue dashed). In all cases we used 100
realisations to compute the average and standard deviation, which
explains the noisy curves. As discussed earlier, we find that
taking the extremum over rotations can increase the detection
power, especially for the $\chi^2$ estimator.

We also see that for the T[4,4,4] topology and the correct orientation,
the correlation method eventually overtakes the likelihood method.
This is most likely because the T[4,4,4] correlation matrix is closer
to being diagonal than the T[2,2,2] correlation matrix. At high $\ell$
the diagonal elements start to dominate the contributions to the
$\chi^2$. The correlator method is not sensitive to this contribution
as it does not sum over the diagonal elements. After maximising over
orientations, on the other hand, the likelihood is always superior
to the correlation method, except maybe for the highest $\lmax$.

We further notice that the detection power keeps increasing with
increasing $\lmax$, even though things tend to slow down beyond
$\ell\approx40$. This means that it is useful to consider the
largest $\ell$ for which we have the correlation matrix and
which we can analyse in a reasonable amount of time. Unfortunately, it is also
the case (and hardly surprising) that the smallest universes
profit the most from analysing smaller scales. The traces from
large but finite universes become rapidly weaker as $\lmax$ 
increases. As there is little practical difference between
a 20 $\sigma$ detection and a 50 $\sigma$ detection, it seems
in general quite sufficient to consider scales up to $\lmax=40$
to $60$. The higher $\ell$ may become more important when we
also consider the ISW effect.

\subsection{What size of the universe can be detected?}

From the suppression of the low-$\ell$ modes in the angular
power spectrum, the T[4,4,4] topology seems a good candidate for
the global shape of the universe. Can we constrain it with
one of our methods? Tables \ref{tab:lambdarot} and
\ref{tab:chi2rot} show that we can indeed distinguish a universe
with T[4,4,4] topology from an infinite one at over 10 $\sigma$.

\begin{figure}[ht]
\begin{center}
\includegraphics[width=70mm]{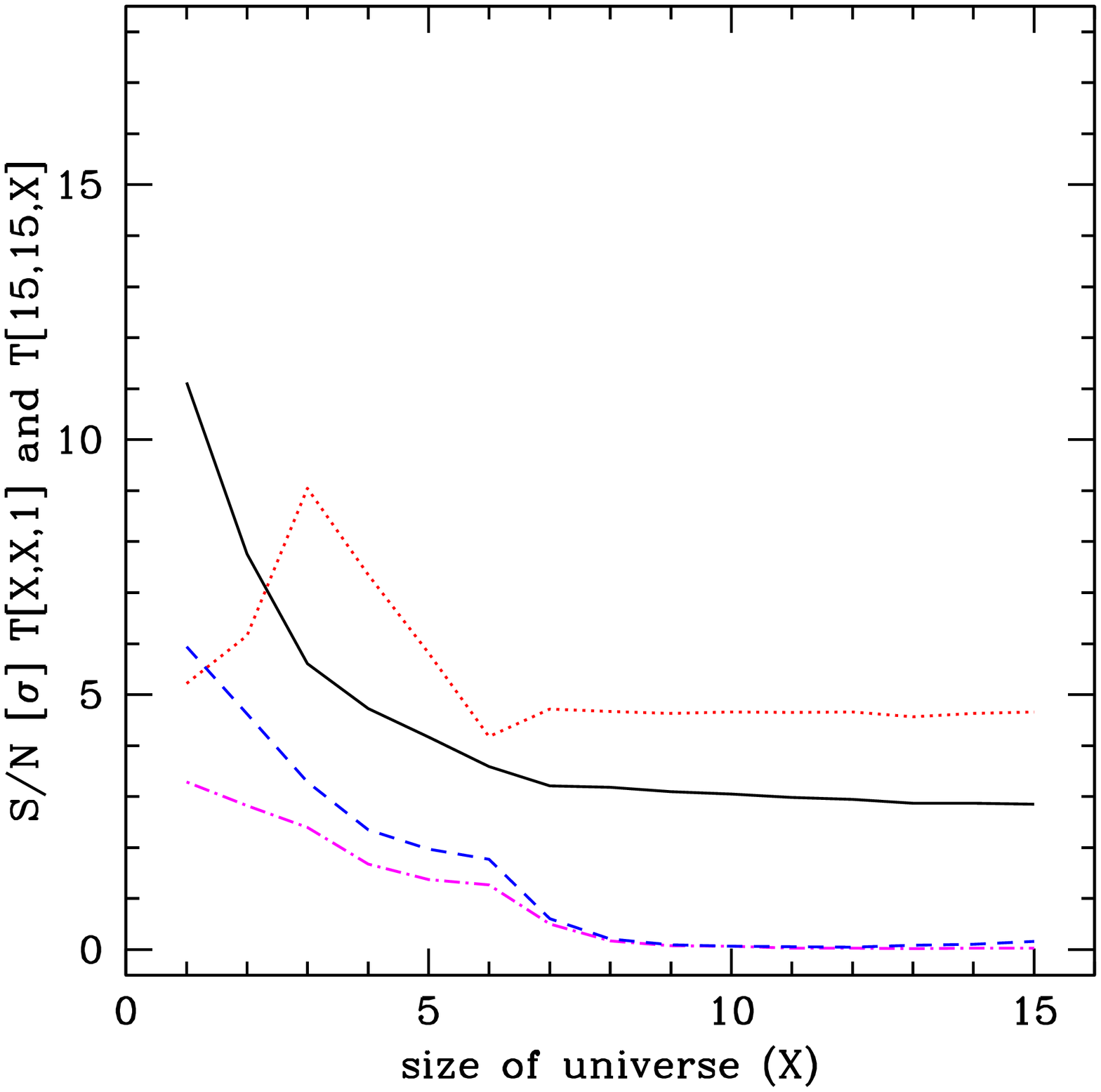} \\
\caption{\label{fig:Xscal_norot} 
Detection significance assuming that we know the correct
orientation. The topologies were T[X,X,1] (solid black
and dotted red line) and T[15,15,X] (dashed blue and dot-dashed magenta line).
The estimators were respectively the correlation amplitude $\lambda$
(dotted red and dot-dashed magenta line) and the likelihood $\chi^2$
(solid black and dashed blue line). We used $\lmax=16$.} 
\end{center}
\end{figure}

\begin{figure}[ht]
\begin{center}
\includegraphics[width=70mm]{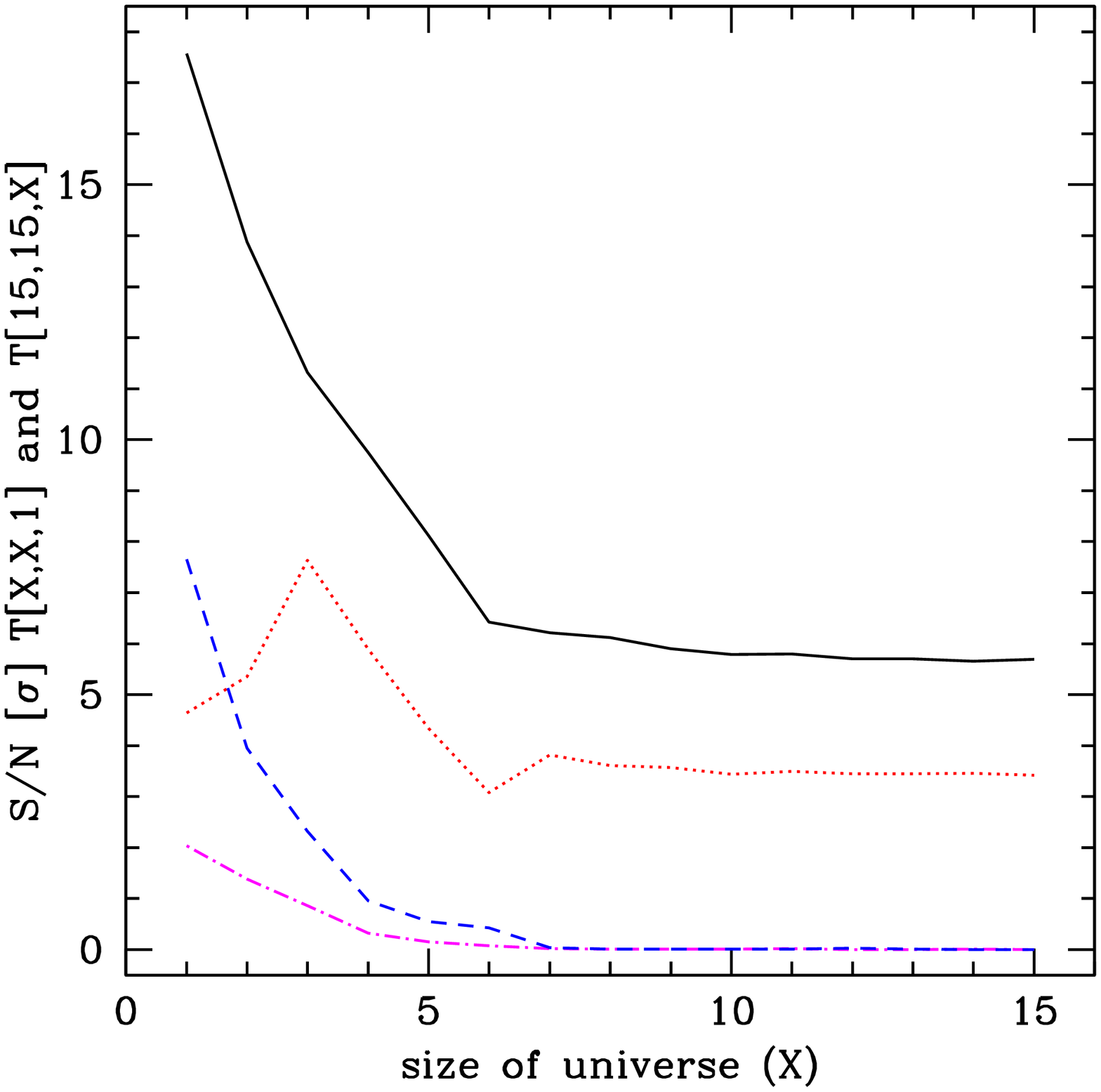} \\
\caption{\label{fig:Xscal_withrot} Detection significance when maximising
over all orientations. The topologies were T[X,X,1] (solid black
and dotted red line) and T[15,15,X] (dashed blue and dot-dashed magenta line).
The estimators were respectively the correlation amplitude $\lambda$
(dotted red and dot-dashed magenta line) and the likelihood $\chi^2$
(solid black and dashed blue line). Again $\lmax=16$.}
\end{center}
\end{figure}

As in the previous section we plot in Figs.~\ref{fig:Xscal_norot} and
\ref{fig:Xscal_withrot} the detection significance
both before and after maximising over directions. This time
we study two families of slab spaces. The first one, T[X,X,1], has one
very small direction of size $1/H_0$ and we vary the other two. We find that we
can clearly detect this kind of topology at $\lmax=16$ for any size
of the larger dimensions. For this example-topology it is very
striking how the correlation estimator is better if we use the
``correct'' alignment, while the $\chi^2$ becomes more powerful
as we extremise over orientations.

The second family, T[15,15,X] is considerably harder to detect as
here two directions are very large and effectively infinite. 
For large values of X we cannot find a difference to an infinite
universe. As the third direction shrinks, we start to see
differences, but only for $X\leq3/H_0$ can we detect the non-trivial
topology at over 2 $\sigma$. In this case the correlation method
is always inferior to the $\chi^2$. In appendix
\ref{app:dkl} we consider a more fundamental distance measure
between correlation matrices, namely the Kullback-Leibler
  divergence.  We confirm that we will never be
able to distinguish T[15,15,X] with $X>6/H_0$ from an infinite universe,
see also Fig.~\ref{fig:dkl_T[15,15,X]}. This is not very surprising,
as in this case the universe is in all directions larger than the
particle horizon today.

\section{A simplified analysis of WMAP data\label{sec:wmap}}

To illustrate the application of these tests to real data, we
perform a simplified analysis of the WMAP \cite{wmap} data. Simplified
in the sense that we do not deal with issues like map noise and sky cuts.
In general, one has to simulate a large number of maps where
both of these effects are included, and which are then analysed with the
same pipeline as the actual data map. However, as an illustration we
will analyse reconstructed full-sky maps. We use the
internal linear combination (ILC) map created by the WMAP team, which
we will call the WMAP map from now on. We also use two map reconstructions
by by Tegmark, a Wiener filtered map (TW) and
a foreground-cleaned map (TC) \cite{tegmap}. All of these maps are
publicly available in HEALPix format \cite{healpix} with a resolution
of $N_{\rm side}=512$. We use this software package to read the map files
and to convert them into $\alm$.

\begin{figure}[ht]
\begin{center}
\includegraphics[width=70mm]{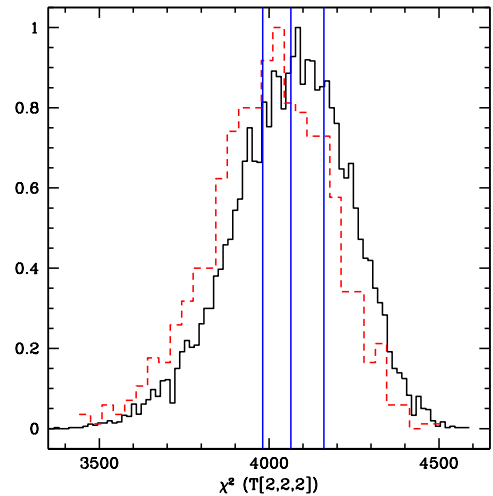} \\
\caption{\label{fig:lilcdist} The distribution of the
$\langle\chi^2\rangle_\infty$ estimator values when testing for
a T[2,2,2] universe with $\lmax=16$. The black solid histogram is computed from
10000 noiseless full-sky realisations used throughout this paper, while
the red dashed histogram used 1000 simulated LILC maps (see text). The vertical 
lines show the $\chi^2$ values of the measured maps, from the left LILC, TW and WMAP
(coincident) and TC.}
\end{center}
\end{figure}

To get some idea of the systematic errors in this analysis, we additionally
analyse the ILC map reconstructed by Eriksen \etal~(LILC). They also
produced a set of simulated LILC maps (for the trivial topology) with the
same pipeline \cite{lilc1,lilc2}.  It is a necessary (but not sufficient) condition
to trust our simplified analysis that the results from these maps are
consistent with our results for an infinite universe. As an illustration we plot
in Fig.~\ref{fig:lilcdist} the distribution of $\chi^2$ for our simple infinite 
universe maps (black solid histogram) and for the simulated ILC maps which contain noise
and foreground contributions (red dashed histogram). We see that the two
distributions agree quite well, to within their own variance. The variance
observed between the different reconstructed sky maps (WMAP, TC, TW and LILC)
is of the same order of magnitude. This example is for T[2,2,2] and $\lmax=16$, but
it is representative of the other cases.

For our standard example, the T[4,4,4] template, we find a maximal value for the 
1st year WMAP ILC map of $\lambda_{\rm max} = 0.20$. This is about
expected for an infinite universe. A universe exhibiting a genuine
T[4,4,4] topology should lead to roughly $\lambda_{\rm max} = 1$.

\begin{table}[ht]
\begin{tabular}{|lc|ccc|ccc|}
\hline
topology & $\lmax$ & $\chi^2$ & $P_\infty$ & $P_\BB$ & $\lambda$ & $P_\infty$ & $P_\BB$ \\
\hline
T[2,2,2]     &  60     & 33130    & $0.39$ & $0$        &  0.087    &  $0.20$    & $0$ \\
T[4,4,4]     &  60     & 11020    & $0.40$ & $0$        &  0.20     &  $0.64$    & $0$ \\
T[6,6,1]     &  16     &  8805    & $0.85$ & $10^{-6}$  &  0.37     &  $0.29$    & $10^{-5}$ \\
T[15,15,6]  &  16     &   290    & $0.95$ & $0.01$     &  1.6      &  $0.16$    & $0.84$ \\
\hline
\end{tabular}
\caption{The value of $\chi^2$ and $\lambda$ obtained for the WMAP map,
together with the probability of measuring such a value if the universe
is infinite ($P_\infty$) and if the universe has indeed the topology
that we test for ($P_\BB$). 
\label{tab:ilc}}
\end{table}

We give in table \ref{tab:ilc} the values of $\chi^2$ and $\lambda$
for the WMAP map. The values for the other maps are not very different.
 We also give two probabilities for both estimators,
$P_\infty$ and $P_\BB$. The first one is the probability of measuring
a larger value of $\lambda$ (or a smaller value of $\chi^2$) if the
universe is infinite. $P_\BB$ on the other hand is the probability
of measuring a smaller value of $\lambda$ (or a larger value of
$\chi^2$ if the universe has indeed the topology that we tested for.
For a non-detection of any topology we require $P_\infty$ to be not
too small. A positive detection of a topology on the other hand
requires a larger $P_\BB$. If both probabilities are large then
we cannot detect that topology (as exemplified eg. for the case
of T[15,15,6]).
We compute these probabilities with the best-fitting theoretical PDF,
as discussed in the appendix \ref{app:rot}.

\begin{figure}[ht]
\begin{center}
\includegraphics[width=70mm]{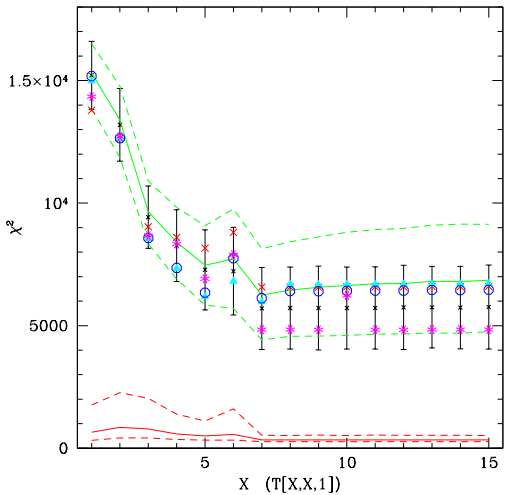} \\
\caption{\label{fig:TXX1} Median and 95\% confidence limits as measured
with the $\chi^2$ estimator for infinite universes (upper green limits) and 
universes with a T[X,X,1] topology (red lower limits), as a function of size
$X$ in units $1/H_0$. We also plot the $\chi^2$
values of the WMAP map (red crosses), the TW map
(cyan triangles) and TC map (blue circles) and the LILC map (magenta stars).
All sky maps are consistent with an infinite universe and not
consistent with a T[X,X,1] topology for any $X$. We also plot errorbars
for the LILC map simulations.}
\end{center}
\end{figure}

\begin{figure}[ht]
\begin{center}
\includegraphics[width=70mm]{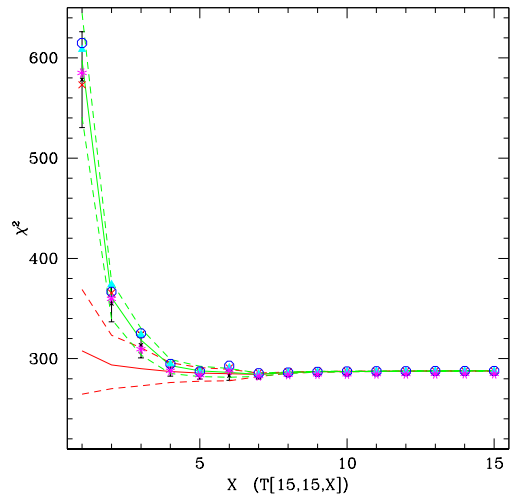} \\
\caption{\label{fig:T1515X} The same as Fig.~\ref{fig:TXX1} for the
T[15,15,X] topology. Again all WMAP maps are consistent with an infinite
universe, but we can only rule out the universes with $X<3$ at more
than 95\% CL.}
\end{center}
\end{figure}

Fig.~\ref{fig:TXX1} shows 95\% confidence limits (estimated numerically
from $10^4$ samples) when testing for the presence (red, lower band) or absence
(green, upper band) of a T[X,X,1] topology. The WMAP data (points) are all compatible with
the infinite universe and rule out this kind of topology very strongly.
The bounds from the simulated LILC maps (black error bars) are consistent with
our simulated maps with a trivial topology, but systematically a bit lower.
We plot the same in Fig.~\ref{fig:T1515X} for a T[15,15,X] topology. Again,
WMAP is compatible with the infinite universe. But as discussed before,
we cannot detect these universes for $X>3/H_0$.
Overall, all results are consistent with an infinite universe.

\section{Bayesian model selection\label{sec:evidence}}

The likelihood can also be used in a purely Bayesian approach.
We are interested in the probability of a model given the data,
$p(\MM|d)$. 
If all topologies are taken to be equally probable, then through Bayes
theorem the statistical evidence $\EE(\MM)$ for a model is proportional to the 
probability of that model, given the data. 
Using the three Euler angles as parameters $\Theta$,
defining the model $\MM$ to be a given topology, and the data $d$
the measured $\alm$ we can write the model evidence as
\be
\EE(\MM) \propto p(d|\MM) = \int \mu(\Theta) \pi(\Theta) \LL(\Theta) ,
\ee
where $\pi(\Theta)$ is the prior on the orientation of the map, see
eg.~\cite{mckay}.
The ratio of the evidence for two topologies
is a Bayesian measure of the relative probability. We
can think of it as the relative odds of the two topologies.
A similar method to constrain the topology was applied previously to
the COBE data, see \cite{inoue}. 

The measure $\mu(\Theta)$ on SO(3)
needs to be independent of the orientation\footnote{The prior and the measure
play a similar r\^{o}le and could be combined into a single quantity.
We prefer to keep them separate here to avoid confusion.}, which pretty much singles
out the Haar measure (up to an irrelevant constant). In terms of the 
Euler angles it is
$d\alpha d\beta d\gamma \sin(\beta)/(8\pi^2)$ with $\alpha$ and $\gamma$
going from $0$ to $2\pi$, and $\beta$ from $0$ to $\pi$. The volume
of SO(3) is then $\int \mu(\Theta)=1$. A simple way to generate random
orientations is to select $\alpha$ and $\gamma$ uniformly in $[0,2\pi]$
and $u$ in $[-1,1]$ and then set $\beta=\arccos(u)$.

The advantage of using Bayesian evidence is that it provides a
natural
probabilistic interpretation which depends only on the actually
observed data, but not on simulated data sets.
Because of this, there is no need to run large
comparison sets. This is a very different view point from
the frequentist approach followed so far.

For an infinite universe the correlation matrix is diagonal and
rotationally invariant (due to isotropy). The integral over the
alignment becomes trivial in this case. If we use whitening then
the correlation matrix is just the unit matrix and we have
\be
\chi^2 = \sum_s |a_s|^2 = \smax.
\ee
The second equality is due to the whitening. The likelihood is then
\be
\LL_\infty = \frac{{\rm const}}{|1|^{1/2}} e^{-1/2\chi^2(\theta)}
= {\rm const}\, e^{-\smax/2} ,
\ee
where the constant normalisation is independent of the topology. We will
neglect it as it drops out when comparing the evidence for different
models. This ``infinite'' evidence gives us a reference point, with
our choice of measure on SO(3) and of normalisation it is
\be
-\log(\EE_\infty) = \frac{\smax}{2}.
\ee

On the other hand, if the universe is infinite then
we know that the expected $\chi^2$ is the trace of the inverse of the 
correlation matrix that we test for. It is again rotationally invariant
as $\langle a_s a_{s'}^*\rangle$ is rotationally invariant. The log-evidence
is on average
\be
-\log(\EE) = \frac{1}{2}\left(\tr(\BB^{-1})+\log|\BB|\right).
\ee
We notice that the expected log-evidence difference to the true
infinite universe is the Kullback-Leibler divergence,
\be
\Delta \log(\EE) = D_{KL}(1||\BB) 
= \frac{1}{2} \left(\log|\BB|+\tr(\BB^{-1}-1)\right)
\label{eq:infev}
\ee
We should not forget though that this is a very crude approximation to
the evidence.
Nonetheless, Eq.~(\ref{eq:infev}) gives a useful indication of the odds
that we can detect a given topology, as it can be evaluated very rapidly,
without performing the integration over orientations. Fundamentally,
this is the amount of additional information about topology contained
in the correlation matrix $\BB$. If the amount of information is not
sufficient to distinguish it from an infinite universe, no test will
ever be able to tell the two apart.
We discuss the Kullback-Leibler (KL) divergence and its possible
applications in more detail in appendix \ref{app:dkl}.

\begin{figure}[ht]
\begin{center}
\includegraphics[width=70mm]{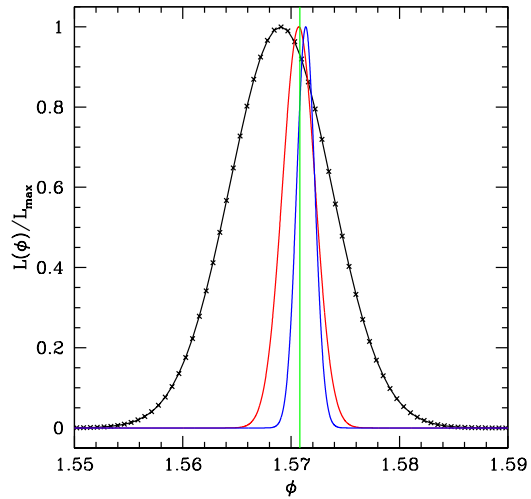} \\
\caption{\label{fig:like} Relative likelihood for a T[4,4,4] topology around
one of the symmetry points where a simulated T[4,4,4] map aligns correctly.
The broadest (black) curve is for $\lmax=16$, the intermediate (red) curve  
$\lmax=32$ and the narrowest (blue) curve $\lmax=60$.
The vertical green line lies at $\phi=\pi/2$. The crosses show the location of
$10^4$ points between $\phi=0$ and $\phi=2\pi$. }
\end{center}
\end{figure}

Of course, faced with real data we have to evaluate the actual
evidence integral.
Unfortunately the likelihood is extremely strongly peaked around
the correct alignments (especially for a non-trivial topology), 
and it is very difficult to sample from it. Already the $\lambda$
and $\chi^2$ estimators require a very precise alignment to reach
the true maximum or minimum. Exponentiating $-\chi^2$ leads to much
narrower peaks in the extrema, and makes the problem far worse.
In Fig.~\ref{fig:like} we plot the relative likelihood (normalised
to unity at the peak) for a universe with T[4,4,4] topology close to
a correct alignment (the vertical line), and for different $\lmax$. 
The broadest peak corresponds to $\lmax=16$, and we added the
location of $10^4$ points evenly spaced between $0$ and $2\pi$ as
black crosses.
This corresponds to a total of roughly $10^{11}$ points to cover all of
SO(3). For $\lmax=16$ we could get away with using only every 10th
point (about $10^8$ points in total) and still detect the high-likelihood
region. But not so for $\lmax=32$ and $60$ (the narrower peaks), which
would easily be missed.

This renders methods like thermodynamic integration infeasible. On
the other hand, we are dealing only with three parameters. Direct
integration is therefore marginally possible by using an adaptive
algorithm. For $\lmax=16$ we need to start out with at least $10^6$ points 
in order to detect the high-probability regions at all. This means that we 
have to count on $10^7$ to $10^8$ likelihood evaluations. The situation
gets worse for higher resolution maps, as both the likelihood evaluations
require more time and the high-probability regions shrink. We therefore
only quote results for $\lmax=8$ in this section. 

\begin{table}[ht]
\begin{tabular}{|lc|cccc|c|}
\hline
topology & $\lmax$ & WMAP    & TC     & TW     & LILC   & $D_{KL}(1||\BB)$ \\
\hline
$\infty$ &  8      &  $-17$  &  $-17$ &  $-17$ &  $-17$ &    0          \\
T[2,2,2]     &  8      & $-114$  & $-103$ & $-100$ & $-102$ &  172          \\
T[4,4,4]     &  8      &  $-46$  &  $-41$ &  $-47$ &  $-44$ &   64          \\
T[6,6,1]     &  8      & $-526$  & $-\infty$ & $-\infty$ & $-\infty$ & 1733          \\
T[15,15,6]  &  8      &  $-17$  &  $-18$ &  $-18$ &  $-17$ &    1          \\
\hline
\end{tabular}
\caption{
The log-evidence $\log_{10}(\EE)$ for a range of topologies and data maps
(see text). We also quote the KL divergence with respect to an
infinite universe for comparison.
\label{tab:evidence}}
\end{table}

As the data sets which define our likelihood we use the same four maps
as in the frequentist analysis: The ILC map by the WMAP team (WMAP), two
maps by Tegmark et al, the Wiener filtered map (TW) and the foreground-
cleaned map (TC) and the ILC map by Eriksen \etal~(LILC).
We quote the logarithm (to base 10) of the evidence in table \ref{tab:evidence} for
our usual range of example models. The relevant quantity for model comparison is
the difference of these values (corresponding to the ratio of the probability). If the
log evidence of a model $A$ is 3 higher than the log evidence of model $B$, we
conclude that the odds for model $A$ are $10^3$ times better. This can be seen
as fairly good odds in favour of model $A$. We plot in Fig.~\ref{fig:erf} the
correspondence between the logarithm of a probability ratio and the number
of standard deviations ($\sigma$) for a Gaussian random variable.

\begin{figure}[ht]
\begin{center}
\includegraphics[width=70mm]{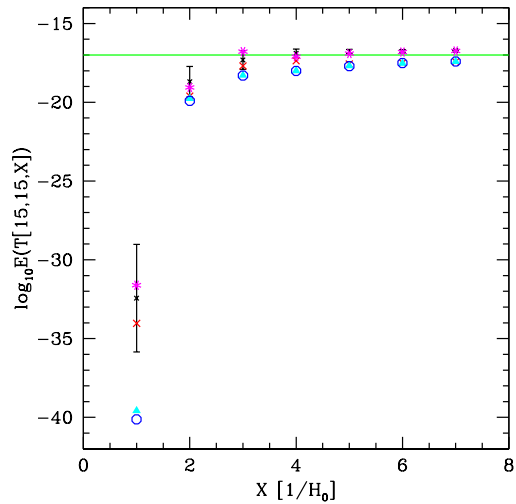} \\
\caption{\label{fig:wmap_ev} The evidence of a T[15,15,X] topology with $\lmax=8$
for four different full-sky reconstructions of the WMAP data 
(WMAP red crosses, TW cyan triangle, TC blue circle and LILC magenta stars).
The black error bars are derived from simulated LILC maps. They are consistent
with the actual LILC data map. The green line shows the predicted evidence of an 
infinite universe.
}
\end{center}
\end{figure}

All topologies except T[15,15,6] are excluded at high confidence. The evidence values
for the different reconstructed CMB maps agree at least qualitatively. We plot
in Fig.~\ref{fig:wmap_ev} the evidence of the T[15,15,X] cases as a function of X.
The two smallest universes are strongly excluded, $X=2$ could
be excluded if we used a higher resolution, and the rest are too close to
the infinite universe to be constrained. We also plot the mean and
standard deviation of the simulated LILC maps as error bars.
 The T[X,X,1] cases are all so completely excluded that the integral
is just barely feasible given the huge numbers involved.

We would like to remind the reader that the results in this section are always
relative to the observed map. It is therefore a bit worrying that the evidences
differ by several orders of magnitude when we consider the different full-sky
reconstructions. We also checked the stability of the results for 1000 simulation 
of the LILC map with known (trivial) topology. We found it to be rather poor
(cf the large error bars in fig~\ref{fig:wmap_ev}), although
this may be partially due to the smaller range of $\ell$. Another possible source
for this lack of stability is our simplistic likelihood. The Bayesian interpretation
of the results is only true if we are able to derive the correct likelihood. This
is an important difference to the frequentist results where we calibrate the
statistical interpretation with the comparison sets. In the frequentist scenario,
we may end up with a sub-optimal test, but we will not get wrong results if we use
the wrong likelihood function. Not so in the Bayesian case, which forces us to
be more careful. A possible way out is to reconstruct a likelihood from the
set of simulated LILC maps.

Normally, a difference
of $2$ to $3$ in $\log_{10}(\EE)$ is taken to be sufficient to strongly disfavour
a model against another one. This may be reasonable for a full analysis that takes
into account all the issues discussed in the following section. For full-sky
reconstructed maps we feel that we should require at least a difference of $10$.
Overall it seems that the frequentist approach leads to results which are more
stable against the uncertainties introduced by the full-sky reconstruction and
foreground removal.

\section{Cosmic Complications\label{sec:complic}}

This paper aims at introducing and discussing the different methods for
constraining the topology of the universe in harmonic space. In doing so
we study an idealised situation with perfect data, neglecting several
issues that are present in the real world. Here we give a quick overview
over the main complications that will have to be dealt with for a rigorous
analysis. Clearly they will change the quantitative results presented
here, but we do not expect that they will lead to qualitative changes in
the results.

\subsection{Noise\label{sec:noise}}

If we assume constant and independent per-pixel noise $\sigma_N$ then
the covariance matrix of the $\alm$ acquires an additional diagonal
term,
\be
\langle a_s^* a_{s'} \rangle + \sigma_N^2 \delta_{ss'} .
\ee
This is fairly close to what many CMB experiments (like WMAP and Planck) 
expect for their data.  The
CMB power spectrum on large scales behaves roughly like $1/\ell^2$ 
(Harrison-Zel'dovich) with a power of about $C_{10}\approx 60 \muK^2$.
For any experiment that probes scales beyond the first peak, we can
conclude that the large scales ($\ell<100$ say) are completely
signal dominated. Taking WMAP as an example, we see that Fig.~1 of 
\cite{wmap_hinshaw} gives a noise contribution
to the $C_\ell$ of $0.1$ to $0.6 \muK^2$ depending on the assembly. As the
noise additionally (to first order) does not enter in the off-diagonal
terms, we can safely neglect it for a first analysis.

More generally we expect a fixed noise variance per detector and
per observation. The resulting
per-pixel noise is $\sigma_N(x) = \sigma_0/\sqrt{N_{\rm obs}}$.
Turning again to WMAP as an example, we find that
they cite a noise variance $\sigma_0 \approx 2 - 7$ mK.
Expressed in terms of the spherical harmonic coefficients, the
correlation matrix in this scenario becomes
\be
\langle a_s^* a_{s'} \rangle + 
\sigma_0^2 \int d^2\!x N_{\rm obs}^{-1}(x) Y^*_s(x) Y_{s'}(x)
\ee
where the integration runs over all pixels $x$. Because of its spatial
variation, the noise is no
longer confined to the diagonal and should strictly speaking be taken 
into account. But the off-diagonal terms will still be very small. 
The most straightforward way to include the noise is to simulate maps with 
the correct power spectrum and noise properties and to co-add them. 
This is especially the case when we deal with a complicated sky cut (see below).

The ILC maps that we used here have more complicated noise properties
due to the full-sky reconstruction. But the noise itself will still be
negligible on large scales, compared to the signal. More worrying are
potential foreground contaminations that were not completely subtracted.
We explore that problem partially in section \ref{sec:wmap} by using
simulated LILC maps.

\subsection{Uncertainties in the cosmological parameters}

So far we have used correlation matrices computed for a fixed cosmological
model. But there are still significant
uncertainties present in the true value of the cosmological parameters,
and even in the underlying cosmological model. An example was recently
discussed in \cite{params}.
In principle we have to take such uncertainties into account. For the 
Bayesian model selection approach, we could do it straight-forwardly by marginalising
over them. Of course this may mean computing a large number of correlation 
matrices for different cosmological models, which would lead to a
computational challenge. Alternatively, one should consider a selection
of models and incorporate the variance of the correlations into a systematic
error on the correlation matrices.

In practise, we hope that the whitening which eliminates differences in the
power spectrum will also minimise the effects due to this parameter
uncertainty. At the very least it will do so for the ``infinite universe''
tests where no off-diagonal correlations are present. The result that
the full-sky WMAP maps are compatible with an infinite universe is thus
not affected by the parameter uncertainty.

\subsection{The integrated Sachs-Wolfe effect \label{sec:isw}}

An issue somewhat related to the last point is that not all perturbations
are generated on the last scattering surface. Some of them are due to the
integrated Sachs-Wolfe (ISW) effect. Especially perturbations due to the
late ISW effect that
are generated relatively close to us are then not affected by the global
topology and carry no information about it. They act as a kind of noise
for our purposes. This contribution is especially problematic when
searching for matching circles in pixel space. It is readily included
when working with the correlation matrices, even though it will also be
subject to the parameter uncertainties and it will lower our
detection power substantially. 

The rapid decrease of the late ISW effect with increasing $\ell$
provides an additional incentive for probing smaller scales,
$\ell\approx40-60$.

\subsection{Sky cuts}

Here we have only considered full-sky maps. Unfortunately a large part
of the sky-sphere is covered by our galaxy which leads
to foregrounds that are not easy to subtract and obscure the true CMB
signal. The most conservative approach is therefore to remove a part
of the sky via a sky cut. This amounts to introducing a mask $\MM(x)$ in
pixel space, with value $1$ on the pixels $x$ where the CMB signal is
clean, and $0$ in the contaminated parts of the sky. We then
consider the pseudo-$\alm$
\be
\hat{a}_{\ell m} = \int d^2\!x \MM(x) \delta T(x) Y_{\ell m} (x)
\ee
instead of the true $\alm$. We can perform the masking operation directly
in harmonic space, using the spherical Fourier transform of the mask,
\be
\MM_{s s'} = \int d^2\!x \MM(x) Y_s(x) Y^*_{s'}(x) .
\ee
The relation between the true $\alm$ and the observed pseudo-$\alm$ is
then given by $\hat{a}_s = \sum_{s'} \MM_{ss'} a_{s'}$. Unfortunately
the mask matrix $\MM$ corresponds to a loss of information and can in
general not be inverted. We could of course use SVD to invert it,
and eliminate the small SVD eigenvalues. However, this would be quite
similar to a full-sky reconstruction. Instead, it may be preferable
to apply the sky cut to the correlation matrix as well. The resulting 
pseudo-correlation matrix is then
\be
\hat{\BB} = \MM^{\dag} \BB \MM .
\ee

This leads to two problems. The first one is purely computational: 
The sky cut has a fixed orientation (with respect to the $\alm$).
So far it did not matter if we rotated the correlation matrix or
the $\alm$, as only the relative orientation counted. But since the
sky cut defines an absolute orientation we now need to apply the
rotation to the correlation matrix. Rotating
the correlation matrix is considerably more costly than rotating the
observed $\alm$. The use of the eigenvector
decomposition (\ref{eq:evec}) and rotation of the effective spherical
harmonics $b_{\ell m}$ can somewhat alleviate the situation if only
a few eigenvalues dominate the sum.

The second problem is that a sky cut and its associated mask matrix
introduce just the kind of correlations between different $\alm$ that
we are looking for. A sky cut will impact significantly on
our ability to constrain large universes. We will have to either
accept this limitation, or hope that better full-sky reconstruction
and component separation methods (for example \cite{comp_sep}) will
become available. However, one would have to demonstrate that such
methods do indeed not change the correlation properties of the $\alm$
in a way that influences the detection of a topology-signature. At
the very least one has to consider such effects as systematic
errors and include them in the error budget of a full analysis.

\section{Conclusions and outlook}

In this paper we have studied three ways to constrain 
the topology of our universe directly with the 
correlation matrix of the $\alm$.
If the primordial fluctuations are Gaussian then
these correlation matrices contain all the information about the
global shape of our universe that is carried by the CMB. By trying
to find their traces in the measured $\alm$ we can construct the
most sensitive probes possible. 

We studied two frequentist estimators, $\lambda$ which describes
the correlation amplitude between the theoretical correlation
matrix $\BB$ and the measured $\alm$, and $\chi^2=a^\dag \BB^{-1} a$.
Although $\lambda$ has certain advantages at high $\ell$ by leaving
out the diagonal terms, we found the $\chi^2$ to be generally superior
after taking into account the random orientation of the observed map. We also computed
the Bayesian evidence, which we found to be a very sensitive probe.
But the angular integration is computationally very intensive, especially
at high resolutions. Additionally, much care is needed in constructing
the likelihood function. For these reasons, 
the $\chi^2$ minimised over rotations seems the most useful of our tests.

For our scenario we find that even high multipoles, $\ell > 50$,
still carry important information about the topology. However,
the amount of work needed to extract the information scales as
a high power of $\ell$. For most cases $\ell \approx 30 - 40$ seems
a sufficient upper limit. 

We finally apply our methods to a set of reconstructed full-sky maps based
on WMAP data. For all topologies considered (cubic and slab tori) we
find no hints of a non-trivial topology. Based on the exclusion of the
T[4,4,4] topology, we conclude that the fundamental domain is at last
$19.3\textrm{Gpc}$ long if it is cubic. We rule out (not very surprisingly)
any universe where a fundamental domain in any direction is smaller
than $4.8$Gpc (based on the T[X,X,1] cases). If the universe is infinite
in two directions, then the third direction has to be larger than
$14.4$Gpc. These limits still allow two copies of the universe inside
the current particle horizon. We prefer to understand this analysis
as a demonstration of our methods, as we neglected a range of important
issues such as the ISW effect.

The noise of the WMAP data is already cosmic variance dominated on
the scales of interest. Future experiments will not be able to provide
significantly better CMB temperature data sets, although some improvement
may come from better foreground separation with more frequencies, and
from e.g. using polarisation maps in addition to the temperature
maps. Short of waiting a few
billion years for the universe to expand further, these tests and especially
the information theoretical limits provided by the Kullback-Leibler divergence
give us an idea about what we can learn of the shape of our universe.

\begin{acknowledgments}

It is a pleasure to thank Andrew Liddle and Peter Wittwer
for helpful comments.
MK and LC thank the IAS Orsay for hospitality; this work
was partially supported by CNES and the Universit\'{e}
Paris-Sud Orsay.
MK acknowledges financial support from the Swiss National
Science Foundation. Part of the numerical analysis was
performed on the University of Geneva Myrinet cluster.

\end{acknowledgments}

\appendix

\section{Finding an optimal estimator\label{app:opt}}

It is interesting to compare the expressions for the $\chi^2$ and the 
$\lambda$ estimator.
The philosophy of the two approaches is very different. In the first
case we write down a likelihood function for a given covariance
matrix. In the second case we correlate the noisy estimated
covariance matrix with a theoretical model. We then use the correlation
amplitude $\lambda$ as a measure of goodness-of-fit. To compare the two
methods, we use the eigen-space expansion (\ref{eq:evec}). As $\BB$ is
hermitian, the eigenvalues are real; if we use the full covariance
matrix, which is positive definite, the eigenvalues are also positive.

Introducing this expansion into the expression for $\chi^2$ we find
\be
\chi^2 = \sum_i \frac{1}{\epsilon^{(i)}} 
\left|\sum_s a_s v_s^{(i)}\right|^2 .
\ee
To compute the same for the correlation amplitude, we use that the
eigenvectors are normalised and orthogonal, 
$\sum_s v_s^{(i)} v_{s}^{(j)*} = \delta_{ij}$. The autocorrelation
is then simply $\sum_{ss'}|\BB_{ss'}|^2 = \sum_i \epsilon^{(i)2}$ and
the correlation amplitude is
\be
\lambda = \frac{\sum_i \epsilon^{(i)} \left|\sum_s a_s v_s^{(i)}\right|^2}{
\sum_i \left(\epsilon^{(i)}\right)^2} .
\ee
If one eigenvalue dominates, then the two expressions
coincide. If all eigenvalues are equal, then $\chi^2 = s_{\rm max}\lambda$.
This happens for an infinite universe if we normalise it by the power
spectrum. 
In both cases the statistical properties are equal.

In the intermediate cases we see that both correspond to a different
weighting of the correlations between the eigenvectors and the $\alm$.
The question arising now is whether we can determine an optimal weighting
that leads to the smallest variance if the $\alm$ are drawn either from
an infinite universe or from one with covariance matrix $\BB$. If the
two requirements are not the same, it is preferable to optimise with 
respect to the infinite universe, as large universes will be close to this
case.

Let us, as an example and guided by the above discussion, 
postulate a general estimator
\be
\alpha = \sum_i \alpha^{(i)} X_i
\ee
where we use the structure that we observed above,
\be
X_i = \left|\sum_s a_s v_s^{(i)}\right|^2.
\ee
The expectation value and the variance of the general estimator are
\bea
\langle\alpha\rangle &=& \sum_i \alpha^{(i)} \langle X_i\rangle \\
\sigma^2 &\equiv& \langle\alpha^2\rangle-\langle\alpha\rangle^2 \nonumber\\
&=& \sum_{i,j} \alpha^{(i)}\alpha^{(j)} 
\left(\langle X_i X_j \rangle-\langle X_i\rangle\langle X_j\rangle \right)
\eea

The aim is to find the $\alpha^{(i)}$ that minimise the variance
of the estimator, subject to a normalisation constraint. We are
going to consider several different limits. The simplest
example is the case where the eigenvectors are due to an infinite
universe, in which case $v_s^{(i)} = \delta_{is}$. It is now easy
to see that $\langle X_i\rangle = \langle a_i a_i^*\rangle=C_i$ and 
$\langle X_i X_j\rangle=C_i C_j + 2 |\AA_{ij}|^2$ where $\AA_{ij}$ is
the covariance matrix from which the observed $\alm$ are drawn.
The expectation value and variance of the general estimator are now
\bea
\langle\alpha\rangle &=& \sum_i \alpha^{(i)} C_i \\
\langle\alpha^2\rangle - \langle\alpha\rangle^2 &=&
2 \sum_{ij} \alpha^{(i)} \alpha^{(j)} |\AA_{ij}|^2
\eea

Adding the constraint $\sum_i \alpha^{(i)} = \smax$ with a Lagrange
multiplier $l$ we have to minimise the expression
\be
2 \sum_{ij} \alpha^{(i)} \alpha^{(j)} |\AA_{ij}|^2 
+ l\left( \sum_i \alpha^{(i)} - \smax \right).
\ee
The relevant system of equations is found as usual by computing the
first derivatives with respect to $l$ and the coefficients and
setting them to zero to find the extrema:
\bea
\sum_i \alpha^{(i)} &=& \smax \\
l + 4 \sum_i \alpha^{(i)} |\AA_{ik}|^2 &=& 0 \quad \forall k = 1,\ldots,\smax
\eea
This is a linear system which can be solved via matrix inversion. 
For the simplest case where $\AA_{ij}=C_i \delta_{ij}$ (the observed
$\alm$ are also drawn from an infinite universe with power spectrum
$C_i$) we can write down the solution up to a normalisation constant:
\be
\alpha^{(i)} \propto \frac{1}{C_i}
\ee
We assume that both the template and sky have the same power
spectrum $C_i$. In our case this also means that the eigenvalues
of $\AA$ are $\epsilon^{(i)} = C_i$. The minimum variance estimator
is therefore proportional to the $\chi^2$. On the other hand, after
whitening $C_i=1$ and both estimators become equivalent.

It is also easy to consider the case when the $a_s$ are distributed
according to the same correlation matrix $\BB$ that we compare them
with. As the eigenvectors are orthonormal, we find that
\bea
\langle X_i \rangle_\BB &=& \epsilon^{(i)} \\
\langle X_i X_j \rangle_\BB &=& \epsilon^{(i)} \epsilon^{(j)} 
       + 2 (\epsilon^{(i)})^2 \delta_{ij}
\eea
The variance of the estimator is then
\be
\sigma_\alpha^2 = 2 \sum_i \alpha_i^2  (\epsilon^{(i)})^2 .
\ee
This is minimised by 
\be
\alpha_i\propto\frac{1}{\epsilon^{(i)}}
\ee
as before. The $\chi^2$ estimator has therefore the minimal variance in
this case as well.

However, we see from tables \ref{tab:chi2} and \ref{tab:chi2div} that
the dominant error is not $\langle\rangle_\BB$ but $\langle\rangle_\infty$.
We should therefore try to minimise this variance instead. Here the
$\alm$ are those of an infinite universe while the eigenvectors are
those of the correlation matrix $\BB$. It is possible to derive an
optimal estimator for this case, but it is rather unwieldy.

Finally, our aim is to maximise the detection of a given
topology. This is not necessarily the same as minimising the variance
as discussed above. Firstly, the discussion necessarily disregards the
deviations introduced by dividing the $\alm$ by their own power
spectrum. Secondly, the $\lambda$ correlation estimator gains power
by leaving out the diagonal terms. And thirdly, we use the extremum
over all orientations which will also change the results.

\section{Extreme value distributions\label{app:rot}}

\begin{figure}[ht]
\begin{center}
\includegraphics[width=70mm]{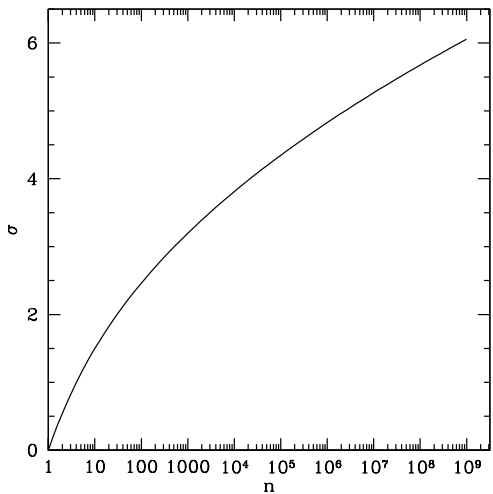} \\
\caption{\label{fig:extreme} The location of the median value in
number of standard deviations $\sigma$ for the maximum value
out of $n$ Gaussian random variables.}
\end{center}
\end{figure}

Computing the extrema of the estimators for a large number of
cases takes a lot of cpu time.
It is important to use this information efficiently, for example by
fitting a theoretically motivated distribution function. We try to
derive such a fitting distribution 
by considering all rotations as independent random realisations. We
then use the maximum or minimum, depending on the estimator. This is 
known as extreme value statistics \cite{ext_val}. For example in the case of $\lambda$
we found that its distribution is nearly Gaussian. We find with
our approximations for the distribution of the maximum out of $n$ draws
\bea
C_n(z) &=&p\left[\text{max}(\lambda_1,\ldots,\lambda_n)\leq z\right] \\
&=& p(\lambda_1\leq z,\ldots,\lambda_n\leq z) \\
&=& \prod_{i=1}^n p(\lambda_i\leq z) \\
&=& C(z)^n .
\eea
Here $C(z)=(1+\text{erf}(\sqrt{2}z))/2$ is the cumulative probability
function of a single univariate Gaussian random variable and $C_n(z)$
the same for the maximum of $n$ independent univariate Gaussians. The 
median lies at $C_n(z)=1/2$ or $C(z)=2^{-1/n}$. We show in figure
\ref{fig:extreme} the location of the median as a function of $n$.
For the relevant number of independent rotations, we find a shift of
$4$ to $6$ sigma. 

A theorem similar to the central limit theorem says that 
there are certain limiting distributions to which the distribution
of an extremal value converges.
The limiting distribution for an unbounded variable like $\lambda$ is the 
Gumbel distribution, with a probability distribution function (PDF)
of the form
\be
P(x)=\exp(-z-\exp(-z))/\sigma \quad {\rm where~} z=(x-\mu)/\sigma
\ee
(see e.g.~\cite{coles} for a discussion and another astrophysical
application).
The expectation value is $\sigma \gamma + \mu$ and the variance is
$\sigma^2\pi^2/6$ where $\gamma\approx0.577$ is the Euler constant.
We can use these two values to find $\sigma$ and $\mu$ given the
variance and expectation value of the distribution.

The cumulative distribution function (CDF) is 
\be
F(x)=e^{-\exp(-z)} ,
\ee
We can consider e.g.~$F(x=0.95)$ as two-sigma upper limit. We find
that for $N$ of the order of a few thousand, $5\sigma$ is a very 
conservative upper bound. Even though the extreme value distribution
moves the expectation values up or down, the variances around those
values can still remain surprisingly small. The signal to noise ratio
need not decrease because of the shift. Indeed, as discussed in \ref{sec:res}
we find that it often even increases.

For a bounded variable like a $\chi^2$ the situation is similar, 
except that the
limiting distribution is now called Weibull distribution, with
\be
P(x)=\frac{\gamma}{x} \left(\frac{x}{\alpha}\right)^\gamma
\exp\left\{-\left(\frac{x}{\alpha}\right)^\gamma\right\} .
\ee
The two parameters $\alpha$ and $\gamma$ can be fixed again by
measuring the expectation value $\mu=\alpha \Gamma[1+1/\gamma]$
and variance $\sigma^2 = \alpha^2(\Gamma[1+2/\gamma]-\Gamma[1+1/\gamma]^2)$ 
of the numerical distribution. The CDF is simply
\be
F(x) = 1-e^{-(t/\alpha)^\gamma} ,
\ee
and $x\geq0$.

\begin{figure}[ht]
\begin{center}
\includegraphics[width=70mm]{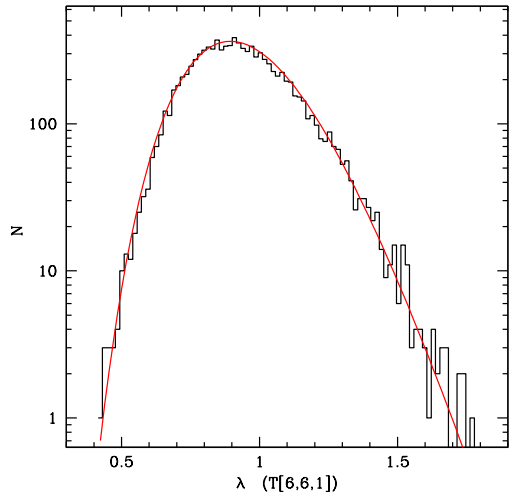} \\
\caption{\label{fig:fit} The PDF of $\lambda$ for a T[6,6,1] topology maximised
over rotations (black histogram, $10^4$ samples) and the best fit using an 
extended Weibull distribution (red curve). The Kolmogorov-Smirnov probability of the
fit is 42\%.
}
\end{center}
\end{figure}

However, we found that this form is
a bad fit even to just the minimum over independent variables
with a true $\chi^2$ type distribution. It seems better to
allow for two different exponents, leading to a PDF of the form
\be
P(x)=\frac{\gamma}{\alpha \Gamma[(1+\beta)/\gamma]} 
\left(\frac{x}{\alpha}\right)^\beta
\exp\left\{-\left(\frac{x}{\alpha}\right)^\gamma\right\} .
\ee
We call this the extended Weibull distribution. The CDF is now 
\be
F(x) = 1 - \frac{\Gamma[(1+\beta)/\gamma,(x/\alpha)^\gamma]}{\Gamma[(1+\beta)/\gamma]}
\ee
where $\Gamma[a,b]$ is the incomplete Gamma function, with
$\Gamma[a,0]=\Gamma[a]$ and $\Gamma[1,x]=\exp(-x)$. We recover
the standard case for $\beta = \gamma-1$. We found the extended
Weibull distribution to be the best-fitting distribution in general,
even for the $\lambda$ estimator. Figure \ref{fig:fit} shows an
example fit to the PDF of $\lambda$ for a T[6,6,1] distribution.

There are different ways to fit the theoretical extreme value distribution
to the numerical CDF. We could for example maximise the Kolmogorov-Smirnov
probability. Instead we decided to use a Bayesian approach: We consider
the numerical values as ``data points'' $d_i$ for the true CDF and use the 
theoretical distribution as a model with parameters $\theta_j$. For each
data point the probability is then given by $p(d_i|\theta)=F(d_i)$. As all
the data points are independent, we can define a likelihood function $\LL$
as
\bea
\chi^2 &\equiv& -2\log(\LL(\theta)) \\
&=& -2 \log \left(\prod_i F(d_i)\right) \\
&=& -2 \sum_i \log \left(F(d_i)\right) .
\eea
We can then easily compute the posterior probability of the parameters $\theta$
that describe the distribution with a Markov-chain Monte Carlo method.

\section{A distance between topologies \label{app:dkl}}

\subsection{The Kullback-Leibler divergence}

Let us consider the following question: What is the expectation value of the
ratio of the likelihoods for covariance matrices $\AA$ and $\BB$
if the $a_s$ are distributed according to a correlation matrix $\AA$?
We have already computed the log-likelihood in section \ref{sec:like}, 
the first case is simply
\be
\langle \log \LL \rangle = -\frac{1}{2} \left(\log|\AA|+ \tr(1)\right) .
\ee
and the second one
\be
\langle \log \LL \rangle = -\frac{1}{2} \left(\log|\BB|+ \tr(\AA\BB^{-1})\right)
\ee
The difference between the two expressions is the logarithm of the 
likelihood ratio,
\be
\langle \Delta\log\LL\rangle = 
-\frac{1}{2} \left(\log|\AA|-\log|\BB| + \tr(1-\BB^{-1}\AA)\right)
\label{eq:kl1}
\ee
This is precisely the Kullback-Leibler divergence between the two
Gaussian distributions described by $\AA$ and $\BB$.

\begin{table}[ht]
\begin{tabular}{|lc|ccc|}
\hline
topology   & $\lmax$ & $\tr(\BB^{-1})$ & $\log(|\BB|)$ & $D_{KL}(1||\BB)$ \\
\hline
$\infty$   &  16     &   288           &      0        &      0           \\
T[2,2,2]   &  16     &  4661           &   -486        &   1944           \\
T[4,4,4]   &  16     &  1570           &   -192        &    545           \\
T[15,15,6] &  16     &   309           &     -8        &      6           \\
T[6,6,1]   &  16     &  20781          &   -399        &   10047          \\
\hline
\end{tabular}
\caption{
Some key quantities for computing the KL divergence. The whitening
enforces $\tr(\BB)=\smax (=288)$.
\label{tab:trdet}}
\end{table}

The Kullback-Leibler (KL) divergence is in general defined for 
two probability distributions $p$ and $q$ as
\be
D_{KL}(p||q) \equiv \int p \log\left(\frac{p}{q}\right) .
\ee
Notice that this is not symmetric, so the symmetrised form
$D(p||q)+D(q||p)$ is sometimes used if it is not clear which distribution
is the fundamental one. In information theory the KL divergence
describes the relative entropy (or information) between the two
probability distributions $p$ and $q$. This corresponds for example
to the amount of information
wasted when trying to describe data distributed as $q$ with a model
based on $p$ (see e.g.~\cite{mckay}).

We consider the KL divergence for random variables $x$ which
have a normal distribution with zero mean and covariance matrix
$\AA$,
\be
p(\AA,x) = (2\pi)^{n/2} |\AA|^{-1/2} \exp\left(-\frac{1}{2} x^T
\AA^{-1} x \right) .
\ee
We can derive an expression for the KL divergence directly in terms of
the covariance matrices by evaluating the Gaussian integrals:
\be
\int p(\AA) \log\frac{p(\AA)}{p(\BB)}
= \frac{1}{2} \left( \log |\BB| - \log |\AA| -
tr\left[ 1-\BB^{-1}\AA\right]\right) .
\ee
This is the same expression as Eq.~(\ref{eq:kl1}).

We have encountered the KL divergence in section \ref{sec:evidence}
where we used it as a zeroth order approximation to the evidence.
In general, it is not rotationally invariant. But although we cannot
use it directly, we can define a distance between two topologies if
their correlation matrices are aligned along the same symmetry axes.
$D_{KL}(\AA||\BB)$ corresponds then to the maximal signal that we
can expect. The base of the logarithm that we use corresponds to a
choice of units -- in information theory the conventional choice
is base 2, corresponding to bits. We quote the numerical results
to base 10, as it makes it easy to interpret the result: 
if $D_{KL}(\AA||\BB)=3$ then we can (at best!) expect to distinguish
the topologies at the 1000:1 level.
If $D_{KL}(\AA||\BB)\leq2$ then it will be very difficult
to distinguish the two topologies. Of course the Kullback-Leibler
divergence depends also on the resolution, $\lmax$.

\begin{figure}[ht]
\begin{center}
\includegraphics[width=70mm]{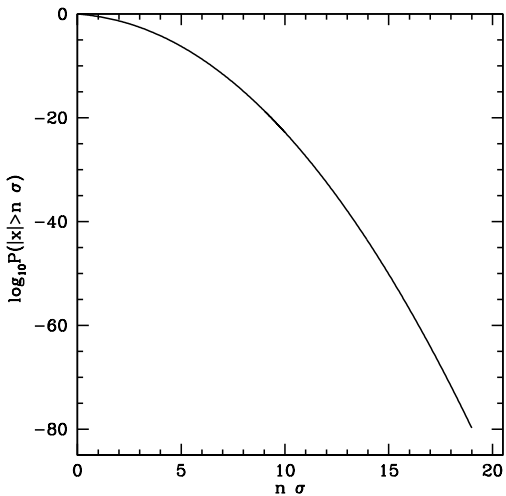} \\
\caption{\label{fig:erf} The probability that a Gaussian random variable
is more than $n$ standard deviations away from the mean. This figure
helps to compare the results expressed in number of $\sigma$ with those
expressed as $\log_{10}(P)$. }
\end{center}
\end{figure}

When comparing to results quoted as number of standard deviations,
we use that for a Gaussian random variable
\be
P(|x|>\nu\sigma) 
= 1-\sqrt{\frac{2}{\pi}} \int_\nu^\infty e^{-x^2/2} dx
= 1-\erf(\nu/\sqrt{2})
\ee
For $x\gg1$ we can well approximate $1-\erf(x)$ by $\exp(-x^2)/(\sqrt{\pi} x)$.
In figure~\ref{fig:erf} we plot $\log_{10}(P(|x|>\nu\sigma))$ against
$\nu\sigma$ to make it easy to compare the two quantities.

\subsection{Information theoretical limits on detecting a topology}

As we have already mention often, a FLRW universe with the trivial topology
is homogeneous and isotropic. Correspondingly its correlation matrix is
rotationally invariant. In this special case also the KL divergence
does not depend on the relative orientation of the two universes.
The quantity $D_{KL}(1||\AA)$ measures therefore directly how much
``information'' separates the universe with the topology described by
$\AA$ from an infinitely large universe. If there is not enough
information, then we will never be able to detect that topology.

Figure \ref{fig:dkl_T[15,15,X]} shows the KL divergence between an infinite
universe and a T[15,15,X] topology for different $X=L/H_0$ and $\lmax$. We see that
the distance falls rapidly for $L>6/H_0$. Even increasing $\lmax$ does not
help as the correlation matrices become essentially identical. Hence, even
though we can still detect correlations in spite of this universe being
larger than the particle horizon in all directions, we will not be able
to distinguish it from an infinite universe at a significant level.
\begin{figure}[ht]
\begin{center}
\includegraphics[width=70mm]{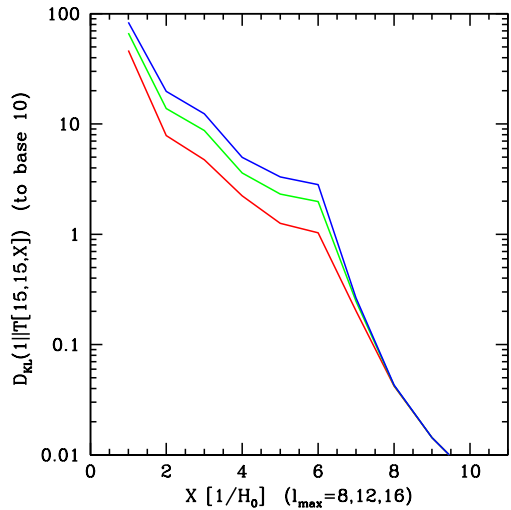} \\
\caption{\label{fig:dkl_T[15,15,X]} The Kullback-Leibler divergence
$D_{KL}(1||T[15,15,X])$ between an infinite universe and a
slab-space, as a function of the size of the smallest dimension $X$.
We show curves for $\lmax=8$, $12$ and $16$. For $X>3$ the topology
becomes difficult to detect and for $X>6$ it is basically impossible
for any $\lmax$. Compare with Fig.~\ref{fig:Xscal_withrot}.}
\end{center}
\end{figure}

\subsection{Comparing different templates}

If the topology of the universe is non-trivial then we will end
up using different correlation matrices until one fits.
If a template is completely wrong we expect
to see no signal at all. 
However, if the template belongs to a topology which is
``similar'' to the real one, then we may find a reduced signal.

\begin{figure}[ht]
\begin{center}
\includegraphics[width=70mm]{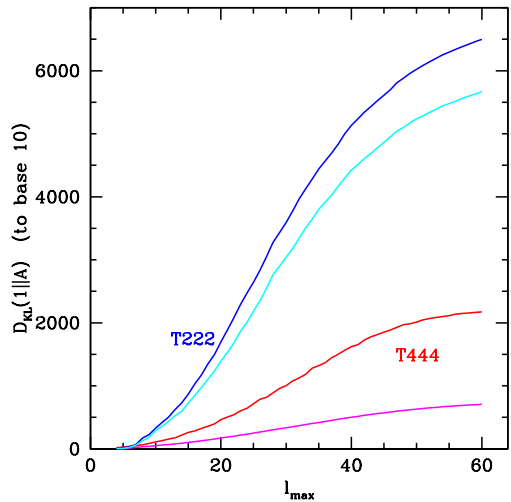} \\
\caption{\label{fig:dkl_TXXX} The scaling of the Kullback-Leibler
divergence as function of $\lmax$. The curves show $D_{KL}(1||T[2,2,2])$
(blue) and $D_{KL}(1||T[4,4,4])$ (red). Both keep increasing for the whole
range of $\lmax$ considered, showing that there is information on
these topologies even at relatively small scales. We also plot
$D_{KL}(T[4,4,4]||T[2,2,2])$ (cyan) and $D_{KL}(T[2,2,2]||T[4,4,4])$. We argue that
the smallness of the latter curve shows that it is possible to detect
a T[2,2,2] universe with a T[4,4,4] template.}
\end{center}
\end{figure}

What does similar mean in this context? As an example, let's assume
that either the universe has a T[2,2,2] topology while we test with T[4,4,4]
or the opposite. In the first case, the signal is actually too strong,
and we end up finding a correlation of order unity
($\lambda=0.91\pm0.05$), but we pay the
price of too much noise. If we had used the T[2,2,2] template, our
detection would have been more significant. On the other hand, if we
use the T[2,2,2] template for a T[4,4,4] universe then the correlation is
smaller ($\lambda=0.11\pm0.02$) while the (non-maximised) value for
infinite universes is $\lambda=0\pm0.02$. Overall, it seems better to test
first the largest universe
that can still be distinguished from an infinite one. 

This is also borne out by the Kullback-Leibler divergence between
T[4,4,4] and T[2,2,2], shown in Fig.~\ref{fig:dkl_TXXX}. We find for example with $\lmax=16$ that
$D_{KL}(T[4,4,4]||T[2,2,2])\approx 2000$ while $D_{KL}(T[2,2,2]||T[4,4,4])=265$.
Both are smaller than $D_{KL}(1||T[2,2,2])$ and the latter is smaller
than $D_{KL}(1||T[4,4,4])$, indicating that it is possible to detect
a $T[2,2,2]$ universe with a $T[4,4,4]$ template. Another possible use
of the Kullback-Leibler divergence is therefore to map out the
space of topologies and to identify those which are very similar.
This helps to reduce the space of models that needs to be searched.

\end{document}